\PassOptionsToPackage{table}{xcolor}
\documentclass[pmlr]{jmlr}

\usepackage[T1]{fontenc}

\RequirePackage{graphicx}
 \usepackage{booktabs}
\usepackage{longtable}
\makeatletter
\def\set@curr@file#1{\def\@curr@file{#1}} 
\makeatother
\usepackage[load-configurations=version-1]{siunitx} 
\usepackage{multirow}
\usepackage{enumitem}
\usepackage{caption}

\theorembodyfont{\upshape}
\theoremheaderfont{\scshape}
\theorempostheader{:}
\theoremsep{\newline}

\jmlrvolume{298}
\jmlryear{2025}
\jmlrworkshop{Machine Learning for Healthcare}

\title[Confidence-Guided Multi-Stage Fusion for Multimodal Data]{\texttt{MedPatch}: Confidence-Guided Multi-Stage Fusion for Multimodal Clinical Data}

\author{\Name{Baraa Al Jorf}
       \Email{baraa.al.jorf@nyu.edu}\\ 
       \addr Engineering Division\\
       \addr NYU Abu Dhabi\\
       \addr Abu Dhabi, UAE\\
       \Name{Farah E. Shamout}
       \Email{farah.shamout@nyu.edu}\\ 
       \addr Engineering Division\\
       \addr NYU Abu Dhabi\\
       \addr Abu Dhabi, UAE\\
       } 

\begin{document}

\maketitle

\begin{abstract}
  Clinical decision-making relies on the integration of information across various data modalities, such as clinical time-series, medical images and textual reports. Compared to other domains, real-world medical data is heterogeneous in nature, limited in size, and sparse due to missing modalities. This significantly limits model performance in clinical prediction tasks. Inspired by clinical workflows, we introduce \texttt{MedPatch}, a multi-stage multimodal fusion architecture, which seamlessly integrates multiple modalities via confidence-guided patching. \texttt{MedPatch} comprises three main components: (i) a multi-stage fusion strategy that leverages joint and late fusion simultaneously, (ii) a missingness-aware module that handles sparse samples with missing modalities, (iii) a joint fusion module that clusters latent token patches based on calibrated unimodal token-level confidence. We evaluated \texttt{MedPatch} using real-world data consisting of clinical time-series data, chest X-ray images, radiology reports, and discharge notes extracted from the MIMIC-IV, MIMIC-CXR, and MIMIC-Notes datasets on two benchmark tasks, namely in-hospital mortality prediction and clinical condition classification. Compared to existing baselines, \texttt{MedPatch} achieves state-of-the-art performance. Our work highlights the effectiveness of confidence-guided multi-stage fusion in addressing the heterogeneity of multimodal data, and establishes new state-of-the-art benchmark results for clinical prediction tasks. 
\end{abstract}

\section{Introduction}
Deep neural networks can combine information from multiple data modalities using various fusion mechanisms, primarily classified as early, joint, and late fusion. Late fusion aggregates information on the prediction-level, whereas early and joint fusion approaches integrate information in the latent space \citep{huang_fusion_2020}. The main difference between the latter two is that they fuse information at different stages and often adopt varying pre-training strategies. Recent work in the medical domain predominantly focuses on joint fusion due to its ability to capture complex interactions across heterogeneous data modalities \citep{hayat_medfuse_2022, khader_medical_2023, krones_review_2025}.

Despite recent advances, significant challenges persist in developing fusion approaches that cater to clinical prediction tasks. One notable trend among existing work is that they focus on a single learning paradigm, and do not leverage the inherent advantages of joint and late fusion simultaneously. For example, a recent study proposed the use of an LSTM-based fusion module for clinical prediction tasks involving chest X-ray images and clinical time-series data \citep{hayat_medfuse_2022}, and another incorporated a transformer-based fusion architecture \citep{khader_medical_2023}. We hypothesize that a \textit{multi-stage} solution that leverages the advantages of both learning paradigms would be more ideal, since healthcare practitioners intrinsically synthesize both unimodal and multimodal insights to obtain a holistic understanding of patient health \citep{munro_documentation_2024}. This is further supported by recent findings \citep{xu_mufasa_2021}, whereby a neural architecture search selected an architecture that incorporated varied stages of fusion. Additionally, medical data frequently exhibits sparsity and is limited in size compared to natural domains, and thus requires careful consideration. Hence, there is a need for flexible approaches that cater to medical data sparsity and heterogeneity.


In this work, we introduce \texttt{MedPatch}, a new multi-stage fusion architecture inspired by the iterative nature of real-world clinical workflows. \texttt{MedPatch} leverages confidence-based token-level patching, based on recent work highlighting the benefits of dynamic token pooling in transformers \citep{pagnoni_byte_2024, nawrot_efficient_2023}. \texttt{MedPatch} also overcomes the challenge of sparse data, allowing flexible processing of samples with missing modalities during training and inference. In summary, we make the following contributions:
\begin{enumerate}
    \item We propose \texttt{MedPatch}, a new multimodal deep neural network that mimics clinical decision-making through multi-stage fusion. Specifically, \texttt{MedPatch} leverages joint and late fusion simultaneously. The joint fusion module combines unimodal tokens based on  token-level confidence, and incorporates a missingness module that indicates the availability of modalities for a given sample. The late module then combines diverse information based on the stage-specific predictors.
    \item We introduce a new multimodal benchmark dataset consisting of four routinely-collected data modalities: clinical time-series data representing a patient's Electronic Health Records (EHR) extracted from MIMIC-IV \citep{johnson_mimic-iv_2023}, Chest X-Ray images (CXR) from MIMIC-CXR \citep{johnson_mimic-cxr_2019}, as well as Radiology Reports (RR) and Discharge Notes (DN) from MIMIC-Notes \citep{johnson_mimic-iv-note_2023}. To the best of our knowledge, this is the first integration of the four data modalities for clinical prediction tasks using the publicly-available datasets. 
    \item We conduct an empirical evaluation of \texttt{MedPatch} and compare it to state-of-the-art baselines for binary in-hospital mortality prediction and multi-label clinical condition classification. \texttt{MedPatch} achieves performance improvements, in terms of the Area Under the Receiver Operating Characteristic curve (AUROC) and the Area Under the Precision-Recall Curve (AUPRC), highlighting its merit in advancing multimodal fusion for healthcare. We make our code publicly available to support evaluation of competing models and reproducibility: \url{https://github.com/nyuad-cai/MedPatch.git}.
\end{enumerate}





 \subsection*{Generalizable Insights about Machine Learning in the Context of Healthcare}
Our work highlights the benefit of integrating information from key medical data modalities in an iterative manner that mimics clinical decision-making, specifically clinical time-series data, chest X-ray images, radiology reports and discharge notes. It specifically highlights the value of synthesizing information from unimodal and multimodal feature extractors,  since clinicians typically alternate between an overall assessment of a patient and more detailed modality-specific analyses \citep{munro_documentation_2024}. Our results demonstrate that such an approach could lead to  performance improvements in clinical prediction tasks, which in turn has the potential to improve clinical decision support systems and improve patient outcomes. Furthermore, our findings motivate future research on model adaptation techniques that can leverage pre-trained encoders, reducing the computational cost of training models from scratch. Overall, our study highlights the value of multimodal fusion using deep neural networks to improve performance in clinical prediction tasks.

\section{Related Work}

\subsection{Multimodal Learning for Clinical Decision Support}
Healthcare practitioners utilize various data modalities in practice for improved diagnostic accuracy and clinical decision-making~\citep{huang_multimodal_2024}. To this end, multimodal deep learning in healthcare seeks to integrate various data modalities, including clinical time-series data, like vital-sign measurements and laboratory test results, medical images, and clinical reports, to enhance predictive accuracy. In the context of clinical prediction tasks in the Intensive Care Unit (ICU), recent work demonstrated the benefit of combining two main modalities: clinical time-series data extracted from the patient's EHR and CXR images.

For example, \cite{shamout_artificial_2021} established that late fusion with simple averaging is a strong baseline for combining predictions computed independently using the two modalities, EHR and CXR images, for deterioration prediction amongst patients with COVID-19. Late fusion combines predictions of different classifiers applied to the input data modalities, which requires pretraining modality-specific models. Although it is simple, straightforward, and interpretable, it does not capture cross-modal interactions via multimodal representation learning. Joint and early fusion combine intermediate features of different modalities in the latent space. \cite{hayat_medfuse_2022} used an LSTM-based fusion module to process representations of the two modalities. In other studies, the authors proposed transformer-based neural network architectures that fuse the two modalities for predicting in-hospital patient survival~\citep{khader_medical_2023} and clinical condition classification ~\citep{yao_drfuse_2024}. One of the main challenges of intermediate fusion is the need to account for missing modalities during training and inference. Many studies sidestep this issue by focusing on idealized, fully observed data \citep{khader_medical_2023, pham_i-ai_2024, lin_empirical_2021}, or use imputation strategies of varying computational complexity \citep{yao_drfuse_2024, lee_learning_2023}.

Overall, these studies highlight the benefit of integrating clinical time-series data and medical imaging to improve performance of clinical prediction tasks. However, most existing work focuses on a single fusion paradigm, whereas our proposed model \texttt{MedPatch} incorporates a multi-stage approach that mimics the iterative nature of clinical decision-making by combining late and joint fusion strategies to better reflect clinical decision-making, where clinicians first interpret each source of information independently before integrating insights across modalities. Additionally, we explicitly handle missing data to ensure that the model remains effective under both fully and partially observed input scenarios.

Clinical notes, like radiology reports and discharge summaries, constitute a rich source of contextual information in clinical decision-making \citep{demner-fushman_what_2009}. It also includes clinical impressions that cannot be captured easily in structured data. Despite their importance, free-text clinical notes remain underutilized in multimodal research combining EHR and CXR images. A key objective of \texttt{MedPatch} is to bridge this gap by integrating textual information and establishing new benchmark results for two widely studied ICU prediction tasks: in-hospital mortality prediction and clinical condition classification.

Furthermore, typically, each input data modality is encoded using a modality-specific neural network architecture, or encoder. Recently, there has been an emphasis on transfer learning, whereby existing feature extractors, also referred to as encoders, are fine-tuned for specific prediction tasks \citep{khader_medical_2023, deznabi_predicting_2021, lin_empirical_2021, lyu_multimodal_2023, niu_deep_2023}. Such modality-specific encoders are pre-trained for a particular task using large unimodal dataset, considering that multimodal datasets are much smaller in size. In the natural language processing domain, which pertains to processing clinical notes, foundation model encoders act as a backbone for further fine-tuning \citep{wan_med-unic_2023, wu_medklip_2023, eslami_pubmedclip_2023, liu_imitate_2025, huang_multimodal_2024, baliah_exploring_2023}. \texttt{MedPatch} builds upon this work by leveraging transfer learning to fine-tune pre-trained encoders for each of its modalities in the multimodal setting.

\subsection{Token Patching}
Inspired by recent advances in language modeling, we propose joint multimodal fusion via confidence-based patching. Patching techniques have demonstrated significant improvements in efficiency and scalability, particularly within architectures that feature explicit encoding and decoding stages. In these settings, patching is employed to dynamically group tokens or bytes into larger, semantically coherent segments, thereby reducing sequence length and computational cost. For instance, \cite{nawrot_efficient_2023} introduces a dynamic token pooling mechanism that learns to segment and pool tokens into variable-sized groups. This method not only preserves linguistic structure but also enables a more efficient processing pipeline in auto-regressive language models. Similarly, the Byte Latent Transformer aggregates raw bytes dynamically into patches based on entropy-driven criteria, which leads to improved inference efficiency and robust scaling compared to fixed tokenization approaches \citep{pagnoni_byte_2024}.

These patching strategies are intrinsically tied to models with encoding and decoding components, where the auto-regressive nature of the task benefits from a sequential grouping and reconstruction process. In contrast, fusion-based architectures—especially those with classification inference heads—are designed to aggregate heterogeneous inputs into fixed-size, joint representations without a decoding phase. As a result, traditional patching mechanisms, which rely on the generative paradigm, do not directly apply to fusion tasks because the sequential decoding and reconstruction objectives are absent.

Our work addresses this gap by integrating a confidence-based patching approach into a fusion framework. By leveraging dynamic patching to create two distinct joint representations — high confidence and low confidence patches — we enable a more nuanced integration of multimodal information for classification tasks. This adaptation is, to our knowledge, the first instance of applying patching techniques in a fusion-based architecture, thereby extending the benefits of dynamic grouping beyond auto-regressive models.


\begin{figure}[t]
    \centering
    \includegraphics[width=1.0\textwidth]{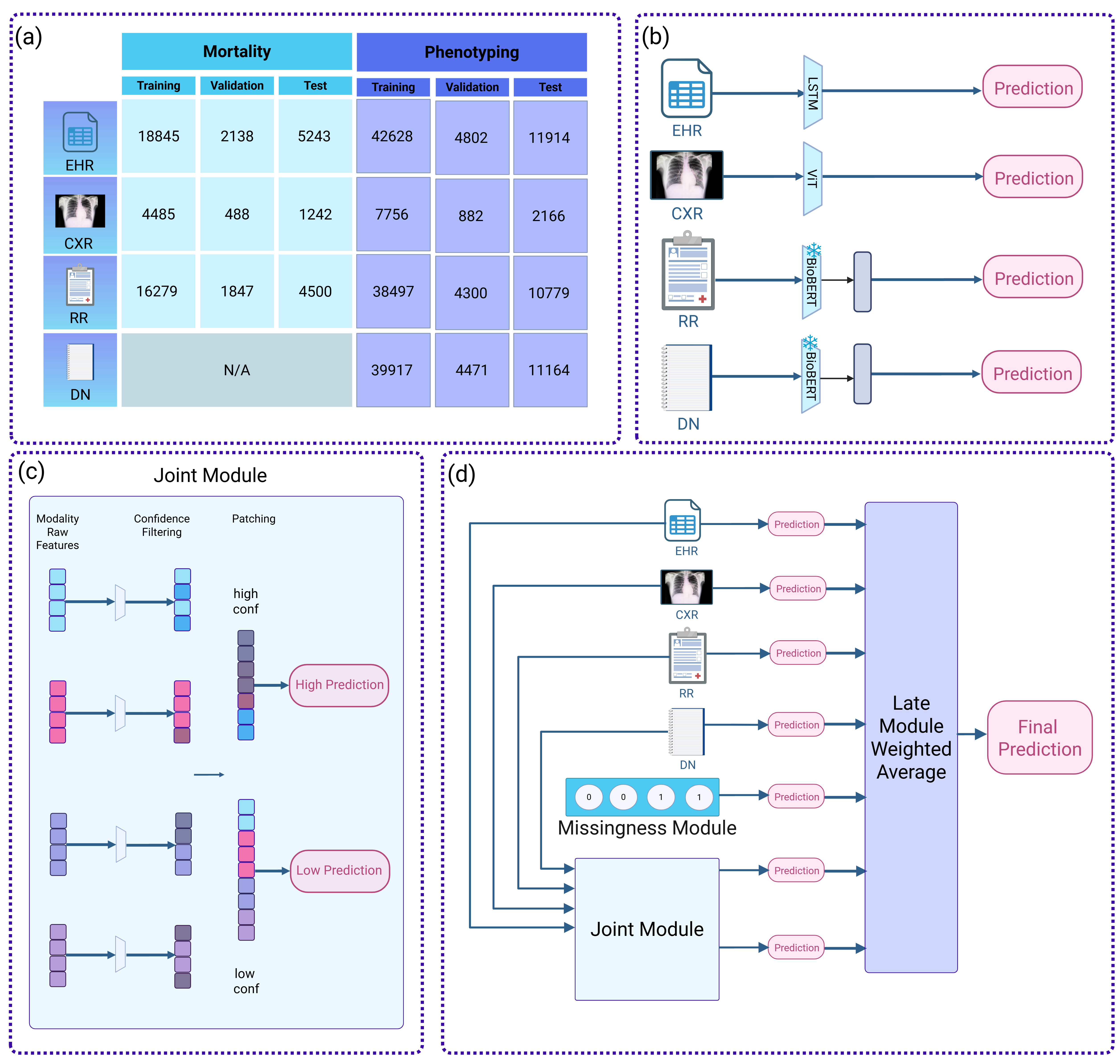} \vspace{-5mm}
    \caption{Overview of the \texttt{MedPatch} framework. (a) Summary of dataset splits. (b) Unimodal pretraining pipeline for each modality. (c) Overview of the joint module used in our architecture. (d) Overview of the MedPatch architecture highlighting all the components including the missingness module, the joint module and its two predictions, and the final prediction returned by the late module. A detailed visualization of the missingness module is presented in Appendix Figure \ref{missing}.}\vspace{-5mm}
    \label{Figure1}
\end{figure}
\section{Methodology}
The overall architecture and training strategy of \texttt{MedPatch} are shown in Figure~\ref{Figure1}. The goal of this study is to enhance model performance in clinical prediction tasks involving four key modalities, denoted as EHR, CXR, RR, and DN, as appropriate for a given task. The main architecture consists of several components described in the next section: pre-trained unimodal encoders, calibrated token-level confidence, confidence-guided joint fusion, explicit missingness indicator, and a late fusion module. 

\subsection{Unimodal Encoders for Feature Extraction}  
Assume a given sample consists of four modalities, specifically $\mathbf{x}=[\mathbf{x}^{(ehr)}, \mathbf{x}^{(cxr)}, \mathbf{x}^{(rr)}, \mathbf{x}^{(dn)}]$, where $\mathbf{x}^{(ehr)} \in \mathbb{R}^{(t\times c)}$ represents structured clinical time-series data of dimensionality $c$ across $t$ time-steps, $\mathbf{x}^{(cxr)} \in \mathbb{R}^{(h\times w)}$ is a CXR image of height $h$ and width $w$, $\mathbf{x}^{(rr)}$ is a sequence of concatenated radiology reports, and $\mathbf{x}^{(rr)}$ is a single sequence representing the patient's discharge note. The goal of \texttt{MedPatch} is to predict a set of ground-truth labels $y$, such that $\hat{y} = \texttt{MedPatch}(\mathbf{x})$. 

For each modality, we define a specific encoder to obtain a tokenized representation and a modality-specific prediction. In particular, for EHR:
 \[
    \mathbf{z}^{(ehr)} = f_{ehr}(\mathbf{x}^{(ehr)}), \quad \hat{y}^{(ehr)} = g_{ehr}(\mathbf{z}^{(ehr)}),
    \]

where $f_{ehr}$ is parameterized as a Long Short-Term Memory (LSTM) network, $g_{ehr}$ is parameterized as a single layer, $\mathbf{z}^{(ehr)} \in \mathbb{R}^{T_{ehr}\times d_{ehr}}$, $T_{ehr}$ is the number of tokens and $d_{ehr}$ is the token embedding dimension. We train the EHR model end-to-end using the full unimodal EHR dataset with the Binary Cross Entropy (BCE) loss.

Similarly for CXR images, 
 \[
    \mathbf{z}^{(cxr)} = f_{cxr}(\mathbf{x}^{(cxr)}), \quad \hat{y}^{(cxr)} = g_{cxr}(\mathbf{z}^{(cxr)} ),
    \]
where $f_{cxr}$ is parameterized as a Vision Transformer (ViT), $g_{cxr}$ is parameterized as a single layer, $\mathbf{z}^{(cxr)} \in \mathbb{R}^{T_{cxr}\times d_{cxr}}$, $T_{cxr}$ is the number of tokens and $d_{cxr}$ is the token embedding dimension. We use an ImageNet pre-trained ViT and train it end-to-end using the full unimodal CXR dataset.

As for RR,
\[
    \mathbf{z}^{(rr)} = f_{rr}(\mathbf{x}^{(rr)}), \quad \hat{y}^{(rr)} = g_{rr}(\mathbf{z}^{(rr)} ),
    \]
where $f_{rr}$ is parameterized as the BioBert model \citep{lee_biobert_2020}, $g_{rr}$ is parameterized as lightweight task-specific linear layers, $\mathbf{z}^{(rr)} \in \mathbb{R}^{T_{rr}\times d_{rr}}$, $T_{rr}$ is the number of tokens and $d_{rr}$ is the token embedding dimension. We freeze $f_{rr}$ and only train $g_{rr}$ using the full RR dataset. Without loss of generality, DN are encoded similarly to obtain $\mathbf{z}^{(dn)}$ and $\hat{y}^{(dn)}$.

    

\subsection{Token-level Confidence Predictors}  


To quantify prediction confidence on the token-level, we introduce confidence predictors $\phi^{(m)}$ for each modality $ m \in \{ehr, cxr, rr, dn\}$ using token-level representations obtained from the pretrained and frozen encoders. Given token embeddings \( \mathbf{z} \) for a given modality $m$, each confidence predictor computes raw logit for the $i$-th token for predicting class $c$:
\[
l_i^{(m,c)} = \phi_i^{(m,c)}({z}_i^{(m)}).
\]

Each predictor is trained by minimizing the BCE with respect to the groundtruth labels, while freezing the main encoders. 



Considering the heterogeneity across the modalities, we further calibrate the confidence scores using temperature scaling. Temperature scaling is applied directly to the raw logits \( l_i^{(m,c)} \). Specifically, the calibrated logit \( \tilde{l}_i^{(m,c)} \) for token \( i \), modality \( m \), and class \( c \) is given by:
\[
\tilde{l}_i^{(m,c)} = \frac{l_i^{(m,c)}}{\tau_i^{(m,c)}},
\]
where \( \tau_i^{(m,c)} \) is a learnable temperature parameter optimized post-training on the validation set. This scaling aligns confidence scores to closely align with true class probabilities.

We then define token-level confidence based on the calibrated logits as:
\[
\gamma_i^{(m,c)} = \max\left(\sigma(\tilde{l}_i^{(m,c)}), 1 - \sigma(\tilde{l}_i^{(m,c)})\right),
\]
where \(\sigma(x) = \frac{1}{1+e^{-x}}\) is the sigmoid function.

\subsection{Joint Fusion via Confidence-based Patching}
After computing the confidence score associated with each token, \texttt{MedPatch} dynamically clusters the tokens of all modalities into two groups: high-confidence and low-confidence modality tokens based on a pre-defined confidence threshold. This results in two pooled representations for a given modality $m$ with varying number of tokens, such that:
\[
{h}_{\text{high}}^{(m,c)} = \rho\{{z}_i^{(m)} \mid \gamma_i^{(m,c)} \geq \theta\}, \quad
{h}_{\text{low}}^{(m,c)} = \rho\{{z}_i^{(m)} \mid \gamma_i^{(m,c)} < \theta\},
\]
where \(\theta\) is a hyperparameter (which we set to 0.75, as 0.5 corresponds to the lowest possible confidence given our definition) and $\rho$ is a pooling function, which we define as simple averaging.

The pooled high-confidence and low-confidence modality representations are processed by a projection layer to obtain $\tilde{{h}}$ and then concatenated. If a modality is missing, we impute zeros for the token vector: 
\[
\tilde{\mathbf{h}}_{\text{high}} = \text{Concat}\left(\tilde{h}_{\text{high}}^{(ehr,c)}, \tilde{h}_{\text{high}}^{(cxr,c)}, \tilde{h}_{\text{high}}^{(rr,c)}, \tilde{h}_{\text{high}}^{(dn,c)}\right),
\]
\[
\tilde{\mathbf{h}}_{\text{low}} = \text{Concat}\left(\tilde{h}_{\text{low}}^{(ehr,c)}, \tilde{h}_{\text{low}}^{(cxr,c)}, \tilde{h}_{\text{low}}^{(rr,c)}, \tilde{h}_{\text{low}}^{(dn,c)}\right).
\]

These concatenated embeddings are passed to separate classifiers:
\[
\hat{y}_{\text{high}} = g_{\text{high}}(\tilde{\mathbf{h}}_{\text{high}}), \quad
\hat{y}_{\text{low}} = g_{\text{low}}(\tilde{\mathbf{h}}_{\text{low}}).
\]

\subsection{Missingness Module}
We define a missingness indicator vector \(\mathbf{a} \in \{0,1\}^4\), where:
\[
a_m = 
\begin{cases}
1, & \text{if modality } m \text{ is available} \\[6pt]
0, & \text{otherwise}
\end{cases}
\]

The Missingness Module processes this indicator vector through a dedicated classifier:
\[
\hat{y}_{\text{miss}} = g_{\text{miss}}(\mathbf{a}).
\]

\subsection{Late Fusion Module}
Finally, we introduce a late fusion module that adaptively combines predictions from all of the architectural components. These include the high-confidence prediction (\(\hat{y}_{\text{high}}\)), low-confidence prediction (\(\hat{y}_{\text{low}}\)), missingness module prediction (\(\hat{y}_{\text{miss}}\)) and the unimodal predictions (\(\hat{y}^{(m)}\)) from each modality \(m\). The predictions are combined using learnable weights \(\alpha_i\), normalized via softmax:
\[
\tilde{\alpha}_j = \frac{\exp(\alpha_j)}{\sum_k \exp(\alpha_k)}, \quad 
\hat{y}_{\text{late}} = \sum_{i}\tilde{\alpha}_k \hat{y}_k,
\]

where \(\hat{y}_k \in \{\hat{y}_{\text{high}}, \hat{y}_{\text{low}}, \hat{y}_{\text{miss}}, \hat{y}^{(ehr)}, \hat{y}^{(cxr)}, \hat{y}^{(rr)}, \hat{y}^{(dn)}\}\). We train the model end-to-end using the following loss function:
\[
\mathcal{L} = \beta_{\text{late}}\,\mathcal{L}_{\text{late}}(y, \hat{y}_{late}) + \beta_{\text{high}}\,\mathcal{L}_{\text{high}}(y, \hat{y}_{high}) + \beta_{\text{low}}\,\mathcal{L}_{\text{low}}(y, \hat{y}_{late}),
\]
using the BCE loss for each loss term.  We allow the model flexibility to dynamically weigh the importance of each loss during training, where the weights are learnable parameters normalized based on a softmax function. 

\section{Experiments} 
\subsection{Clinical Prediction Tasks}
The clinical prediction tasks evaluated in this study encompass tasks typically assessed in recent work pertaining to ICU clinical decision support \citep{yao_drfuse_2024, hayat_medfuse_2022,lyu_multimodal_2023,he_foundation_2025,lin_empirical_2021, yang_multimodal_2021,steinberg_language_2021, liu_attention-based_2023, khader_medical_2023}. These tasks are defined as follows:
\begin{enumerate}[itemsep=1pt]
    \item \textbf{In-hospital Mortality Prediction:} Binary classification task that forecasts whether a patient will succumb during their hospital stay based on information collected within the first 48 hours of ICU admission. We evaluate model performance using the AUROC and the AUPRC. The in-hospital mortality task only considers EHR, CXR, and RR modalities. We purposely exclude DN for this task because it contains information related to patient discharge, which would introduce information leakage.
    \item \textbf{Clinical Conditions Classification:} Multi-label classification task that aims to predict the presence of any of 25 different chronic, mixed, and acute care conditions at the end of an ICU stay. We assess performance using AUROC and AUPRC and use all four modalities as inputs.
\end{enumerate}

\subsection{Dataset Curation and Pre-processing}
We extracted the EHR  data from MIMIC-IV \citep{johnson_mimic-iv_2023}, the CXR images from MIMIC-CXR \citep{johnson_mimic-cxr_2019}, and the RR and DN from MIMIC-IV-Note \citep{johnson_mimic-iv-note_2023}. The MIMIC-IV dataset encompasses a vast array of de-identified health-related data from over $315,460$ patients who stayed in critical care units of the Beth Israel Deaconess Medical Center between 2008 and 2019. It includes comprehensive information such as vital signs, medications, laboratory measurements, observations, and notes, which are essential for robust predictive modeling in healthcare. The CXR images in the MIMIC-CXR Dataset, derived from this cohort, consist of over 377,000 chest radiographs annotated with findings and impressions. The MIMIC-IV-Note dataset supplements this with unstructured textual data, segmented into 331,794 DN and 2,321,355 RR. DN provide summaries of the patient's hospital stay, treatment, and follow-up care, offering insights into patient management and outcomes. RR contain detailed interpretations of various imaging studies, such as CXR, computed tomography, magnetic resonance imaging, and ultrasound. These reports often include comparisons with previous studies and a summarizing impression. Together, these datasets offer a multidimensional view of patient care, facilitating nuanced analyses and predictions through the integration of diverse health data modalities.

We followed the same pre-processing pipeline in previous work to build our benchmarks \citep{hayat_medfuse_2022}. For EHR, we used the same set of 17 clinical variables, five of which are categorical (capillary refill rate, Glasgow coma scale eye opening, Glasgow coma scale motor response, Glasgow coma scale verbal response, and Glasgow coma scale total) and twelve of which are continuous (diastolic blood pressure, fraction of inspired oxygen,
glucose, heart rate, height, mean blood pressure, oxygen saturation, respiratory rate, systolic
blood pressure, temperature, weight, and pH).

For mortality, we discretized the time steps into one hour steps, ending at 48 hours since the first entry. The modalities were paired such that the CXR and RR were recorded within the 48 hours. All radiology reports belonging to a single patient were concatenated as a single sample.  We paired all modalities according to the subject identifier, stay identifier, and hospital admission identifier.
Our pairing strategy mirrors the $\textbf{(EHR + CXR)}_{PARTIAL}$ strategy used in recent work \citep{hayat_medfuse_2022}, where we used all EHR samples and paired any corresponding modalities available, which led to some samples having missing modalities. We split the dataset into training, validation and test sets following previous work \citep{hayat_medfuse_2022, khader_medical_2023}. Figure \ref{Figure1} (a) provides a summary of the dataset size. Supplementary Tables \ref{EHR}, \ref{Prev}, \ref{age}, and \ref{Gender} provide a more detailed description of the data statistics. Supplementary Figures \ref{fig:mortality_diff} and \ref{fig:pheno_dff} provide information on CXR collection times across both tasks.

\subsection{Baseline Models}

To assess how well our model performs we compare it to the following baselines:
\begin{enumerate}[itemsep=1pt]
    \item \textbf{Early Fusion:} This approach fuses modalities at the input level, where the features of each modality, extracted using their respective unimodal feature extractors, are concatenated as a single representation and processed by a classifier \citep{huang_fusion_2020}. The encoders are pretrained in the unimodal setting and then frozen in the multimodal setting.

    \item \textbf{Joint Fusion:} In this baseline, features extracted from each modality-specific encoder are also concatenated and then processed by a classifier. The encoders and classification head are trained end-to-end.

    \item \textbf{Late Fusion:} This fusion technique operates at the output level by aggregating predictions from modality-specific classifiers. Each modality — EHR, CXR, DN, and RR— is processed independently using its respective encoder, and then all of the predictions are averaged to compute the final output.

    \item \textbf{MedFuse:} MedFuse is tailored for processing EHR and CXR using an LSTM module \citep{hayat_medfuse_2022}.

    \item \textbf{MeTra:} MeTra leverages a transformer-based architecture to fuse EHR and CXR features. It processes each modality independently and integrates the outputs through a transformer-based cross-attention mechanism to produce a unified representation. The final prediction is obtained via a classification layer \citep{khader_medical_2023}. 

    \item \textbf{Ensemble models:} Ensemble models are known to boost performance since they aggregate information from multiple experts. We introduce several ensemble models as baselines, specifically a late ensemble, joint ensemble, early ensemble, and diverse ensemble. Each ensemble aggregates the predictions of the top-3 performing models for the respective type of fusion models, while the latter aggregates the top-3 early, late and joint fusion models. 
\end{enumerate}

\subsection{Implementation Details}
We used a single layer LSTM for EHR encoding, ViT small for CXR, and BioBERT for DN and RR.  We note that BioBERT can handle inputs comprising a maximum of 512 tokens. However, since radiology reports and discharge notes often exceed the 512-token limit of the model, we first split each document into non-overlapping chunks of 512 tokens. Each chunk is independently passed through the transformer model which outputs a 768-dimensional embedding for each token. We then apply mean-pooling over the 512 tokens to produce a 512 vector representing the entire document. This final vector is fed into a trainable linear layer for the downstream tasks.

For model training and selection, we conduct extensive hyperparameter tuning experiments using random search. In particular, we fix the batch size to 16, number of epochs to 100, and run 10 sweeps per run for learning rates randomly sampled between $10^{-5}$ to $10^{-3}$ for each model. We implement early stopping with a patience of 15 epochs. The best models are selected based on the validation set AUROC performance. Calibration experiments are run for 5 epochs and we selected the temperature parameters associated with the lowest expected calibration error on the validation set. We report final results on the test sets in terms of AUROC and AUPRC and provide 95\% confidence intervals via bootstrapping. All experiments are run on A100 GPUs using the Adam optimizer.


\section{Results}
In this section, we present our main experimental findings on in-hospital mortality prediction and clinical condition classification, including unimodal, bimodal. and higher-order multimodal performance results as well as ablation experiments that highlight the contributions of individual architectural components. Appendix \ref{tab:auroc_auprc} contains further supplementary analyses offering a detailed comparison of \texttt{MedPatch}'s multiple AUROC and AUPRC calculation approaches. In Appendix \ref{tab:pheno_alpha} we present an extensive breakdown of the $\alpha$ weights assigned by the model for clinical condition prediction per class, and in Appendix \ref{tab:inho_alpha} we present the $\alpha$ weights for in-hospital mortality. These additional analyses help contextualize our results and illustrate how the confidence-guided fusion mechanism and modality-specific weighting contribute to enhanced performance. We summarize the number of trainable parameters per architecture in Appendix \ref{parameters} to highlight the relative efficiency in \texttt{MedPatch}. To further generalize our findings, we experiment with an alternative patching mechanism in Appendix \ref{entropy} and compare our results against a baseline that was not designed for ICU tasks in Appendix \ref{healnet}.
\begin{table}[t]
\caption{Unimodal performance results in terms of AUROC and AUPRC for mortality and clinical conditions classification. We also summarize dataset size for each task, which was further split into training (70\%), validation (10\%) and test (20\%). The text in bold represents the best-performing model.} \vspace{-7mm}
\begin{center}
\resizebox{1.0\linewidth}{!}{%
\begin{tabular}{c c| c c c| c c c}
\toprule
\multicolumn{2}{c|}{\textbf{Unimodal Models}} & \multicolumn{3}{c|}{\textbf{In-hospital Mortality}} & \multicolumn{3}{c}{\textbf{Clinical Conditions}} \\ \midrule 
\textbf{Modality} & \textbf{Encoder} & \textbf{Size} & \textbf{AUROC} & \textbf{AUPRC} & \textbf{Size} & \textbf{AUROC} & \textbf{AUPRC} \\ 
\midrule
EHR & LSTM     & 26226 & \textbf{0.861 (0.847, 0.875)} & \textbf{0.523 (0.484, 0.562)} & 59344 & 0.764 0.752, 0.775)  & 0.423 (0.402, 0.446) \\
CXR & ViT      & 6215 & 0.723 (0.682, 0.760) & 0.351  (0.286, 0.421) & 10804 & 0.692 (0.662, 0.721)  & 0.388 (0.351, 0.431) \\
RR  & BioBERT  & 22626 & 0.742 (0.723, 0.762) & 0.287 (0.258, 0.322) & 53576 & 0.776 (0.764, 0.788) & 0.474 (0.450, 0.499) \\
DN  & BioBERT  & - &   -   &   -   & 55552 & \textbf{0.856 (0.847, 0.865)} & \textbf{0.601 (0.577, 0.625)} \\ 
\bottomrule
\end{tabular}}   \vspace{-5mm}
\end{center}
\label{tab:unimodal}
\end{table}

\begin{table}[t]
\caption{Bimodal performance results in terms of AUROC and AUPRC for \texttt{MedPatch} and baseline models, i.e. using EHR and CXR. The text in bold represents the best-performing model. In Appendix \ref{significane}, we conduct statistical significance testing for the bimodal results using the t-test adjusted for multiple comparisons.}
\begin{center}
\resizebox{0.9\linewidth}{!}{%
\begin{tabular}{lcc|cc}
\toprule
\multirow{2}{*}{\textbf{Models}}  & \multicolumn{2}{c|}{\textbf{In-hospital Mortality}} & \multicolumn{2}{c}{\textbf{Clinical Conditions}} \\ \cmidrule(r){2-5}
& \textbf{AUROC} & \textbf{AUPRC} & \textbf{AUROC} & \textbf{AUPRC} \\ 
\midrule
Early fusion & 0.859 (0.843, 0.873) & 0.522 (0.484, 0.562) & 0.772 (0.760, 0.783) & 0.438 (0.417, 0.462) \\ 
Joint fusion & 0.861 (0.846, 0.875) & 0.526 (0.487, 0.566) & 0.769 (0.758, 0.781) & 0.434 (0.412, 0.457) \\ 
Late fusion & 0.854 (0.838, 0.868) & 0.520 (0.482, 0.558) & 0.582 (0.568, 0.597) & 0.237 (0.225, 0.251) \\ 
MedFuse & 0.861 (0.845, 0.874) & 0.501 (0.462, 0.543) & 0.758  (0.745, 0.770) & 0.418 (0.396, 0.441) \\ 
MeTra & 0.864 (0.850, 0.877) & 0.513 (0.472, 0.552) & 0.766 (0.754, 0.777) & 0.423 (0.401, 0.446) \\ 
\rowcolor[gray]{0.9} \texttt{MedPatch} (Ours) & \textbf{0.868 (0.855, 0.882)} & \textbf{0.541 (0.501, 0.578)} & \textbf{0.773 (0.762, 0.785)} & \textbf{0.439 (0.417, 0.462)} \\ 
\bottomrule
\end{tabular}}   \vspace{-5mm}
\end{center}
\label{tab:bimodal}
\end{table}

\begin{table}[t]
\caption{Model performance across AUROC and AUPRC metrics for mortality prediction (Trimodal: EHR+CXR+RR) and clinical conditions classification (Quatrimodal: EHR+CXR+RR+DN) tasks. The text in bold represents the best-performing model.}
\begin{center}
\resizebox{0.9\linewidth}{!}{%
\begin{tabular}{lcc|cc}
\toprule
\multirow{2}{*}{\textbf{Models}}  & \multicolumn{2}{c|}{\textbf{In-hospital Mortality}} & \multicolumn{2}{c}{\textbf{Clinical Conditions}} \\ \cmidrule(r){2-5}
& \textbf{AUROC} & \textbf{AUPRC} & \textbf{AUROC} & \textbf{AUPRC} \\ 
\midrule
Early fusion & 0.869 (0.853, 0.882) & 0.536 (0.497, 0.576) & 0.772 (0.760, 0.783) & 0.438 (0.416, 0.461) \\ 
Joint fusion & 0.869 (0.854, 0.882) & 0.537 (0.495, 0.576) & 0.771 (0.760, 0.783) & 0.436 (0.414, 0.459) \\ 
Late fusion  & 0.865 (0.850, 0.878) & 0.519 (0.478, 0.558) & 0.825 (0.815, 0.835) & 0.519 (0.497, 0.541) \\ \midrule
Early ensemble & 0.868 (0.854, 0.881) & 0.535 (0.496, 0.572) & 0.754 (0.741, 0.766) & 0.403 (0.383, 0.425) \\ 
Joint ensemble & 0.871 (0.857, 0.884) & 0.545 (0.506, 0.583) & 0.763 (0.751, 0.775) & 0.426 (0.405, 0.450) \\ 
Late ensemble & 0.864 (0.849, 0.877) & 0.528 (0.489, 0.567) & 0.824 (0.813, 0.834) & 0.519 (0.498, 0.541) \\  
Diverse ensemble & 0.872 (0.858, 0.885) & 0.547 (0.509, 0.584) & 0.797 (0.785, 0.808) & 0.474 (0.452, 0.497) \\ 
\midrule 
\rowcolor[gray]{0.9} \texttt{MedPatch} (Ours) & \textbf{0.876 (0.863, 0.890)} & \textbf{0.558 (0.519, 0.597)} & \textbf{0.862 (0.853, 0.871)} & \textbf{0.614 (0.591, 0.638)} \\ 
\bottomrule
\end{tabular}}   \vspace{-3mm}
\end{center}
\label{tab:multimodal}
\end{table}

\begin{table}[t]
\caption{\texttt{MedPatch} Phenotyping subclass analysis performance (AUROC and AUPRC with 95\% confidence intervals) for bimodal and higher-order multimodal setting (i.e. four modalities).}
\centering
\resizebox{\linewidth}{!}{%
\begin{tabular}{lcccc}
\toprule
{\textbf{Phenotype Class}} & \multicolumn{2}{c}{\textbf{Bimodal}} & \multicolumn{2}{c}{\textbf{Multimodal}} \\[2mm]
\cmidrule(lr){2-3} \cmidrule(lr){4-5}
 & \textbf{AUROC (95\% CI)} & \textbf{AUPRC (95\% CI)} & \textbf{AUROC (95\% CI)} & \textbf{AUPRC (95\% CI)} \\
\midrule
Acute and unspecified renal failure           & 0.796 (0.787, 0.805)  & 0.596 (0.577, 0.614)  & \textbf{0.835 (0.827, 0.843)}  & \textbf{0.658 (0.641, 0.676)}  \\[4pt]
Acute cerebrovascular disease                 & 0.907 (0.895, 0.919)  & 0.471 (0.434, 0.512)  & \textbf{0.953 (0.946, 0.959)}  & \textbf{0.647 (0.610, 0.683)}  \\[4pt]
Acute myocardial infarction                   & 0.767 (0.754, 0.781)  & 0.224 (0.202, 0.252)  & \textbf{0.845 (0.833, 0.858)}  & \textbf{0.362 (0.333, 0.396)}  \\[4pt]
Cardiac dysrhythmias                          & 0.694 (0.685, 0.704)  & 0.506 (0.490, 0.521)  & \textbf{0.727 (0.718, 0.737)}  & \textbf{0.553 (0.537, 0.571)}  \\[4pt]
Chronic kidney disease                        & 0.755 (0.744, 0.765)  & 0.459 (0.440, 0.479)  & \textbf{0.807 (0.797, 0.817)}  & \textbf{0.532 (0.511, 0.553)}  \\[4pt]
Chronic obstructive pulmonary disease         & 0.716 (0.703, 0.728)  & 0.308 (0.288, 0.330)  & \textbf{0.759 (0.747, 0.770)}  & \textbf{0.365 (0.343, 0.388)}  \\[4pt]
Complications of surgical procedures or medical care 
                                               & 0.734 (0.723, 0.746)  & 0.407 (0.387, 0.428)  & \textbf{0.769 (0.759, 0.780)}  & \textbf{0.447 (0.425, 0.470)}  \\[4pt]
Conduction disorders                         & 0.757 (0.742, 0.773)  & 0.379 (0.350, 0.408)  & \textbf{0.826 (0.812, 0.839)}  & \textbf{0.494 (0.463, 0.525)}  \\[4pt]
Congestive heart failure; nonhypertensive      & 0.780 (0.772, 0.790)  & 0.556 (0.540, 0.574)  & \textbf{0.847 (0.839, 0.855)}  & \textbf{0.686 (0.669, 0.704)}  \\[4pt]
Coronary atherosclerosis and other heart disease 
                                               & 0.772 (0.763, 0.781)  & 0.608 (0.592, 0.626)  & \textbf{0.811 (0.804, 0.819)}  & \textbf{0.667 (0.652, 0.683)}  \\[4pt]
Diabetes mellitus with complications          & 0.899 (0.891, 0.907)  & 0.592 (0.566, 0.618)  & \textbf{0.904 (0.896, 0.912)}  & \textbf{0.602 (0.576, 0.630)}  \\[4pt]
Diabetes mellitus without complication        & 0.788 (0.778, 0.799)  & 0.409 (0.390, 0.433)  & \textbf{0.819(0.828, 0.809)}   & \textbf{0.459 (0.483, 0.437)}   \\[4pt]
Disorders of lipid metabolism                   & 0.706 (0.697, 0.715)  & 0.616 (0.600, 0.631)  & \textbf{0.782(0.790, 0.773)}   & \textbf{0.699 (0.713, 0.687)}   \\[4pt]
Essential hypertension                         & 0.677 (0.668, 0.686)  & 0.583 (0.569, 0.598)  & \textbf{0.752(0.761, 0.742)}   & \textbf{0.665 (0.681, 0.650)}   \\[4pt]
Fluid and electrolyte disorders                & 0.762 (0.753, 0.770)  & 0.653 (0.638, 0.668)  & \textbf{0.811(0.819, 0.804)}   & \textbf{0.726 (0.739, 0.712)}   \\[4pt]
Gastrointestinal hemorrhage                    & 0.776 (0.762, 0.791)  & 0.220 (0.197, 0.244)  & \textbf{0.919(0.929, 0.909)}   & \textbf{0.597 (0.634, 0.560)}   \\[4pt]
Hypertension 
                                               & 0.746 (0.735, 0.756)  & 0.446 (0.427, 0.467)  & \textbf{0.863(0.871, 0.855)}   & \textbf{0.656 (0.677, 0.638)}   \\[4pt]
Other liver diseases                           & 0.737 (0.723, 0.749)  & 0.310 (0.288, 0.334)  & \textbf{0.872(0.882, 0.863)}   & \textbf{0.613 (0.638, 0.589)}   \\[4pt]
Other lower respiratory disease                & 0.653 (0.637, 0.669)  & 0.172 (0.157, 0.191)  & \textbf{0.718(0.733, 0.704)}   & \textbf{0.226 (0.248, 0.207)}   \\[4pt]
Other upper respiratory disease                & 0.744 (0.720, 0.767)  & 0.236 (0.203, 0.275)  & \textbf{0.836(0.857, 0.817)}   & \textbf{0.377 (0.425, 0.337)}   \\[4pt]
Pleurisy; pneumothorax; pulmonary collapse       & 0.729 (0.712, 0.746)  & 0.172 (0.154, 0.194)  & \textbf{0.847(0.861, 0.833)}   & \textbf{0.388 (0.422, 0.354)}   \\[4pt]
Pneumonia 
                                               & 0.820 (0.808, 0.830)  & 0.396 (0.370, 0.421)  & \textbf{0.902(0.909, 0.895)}   & \textbf{0.564 (0.594, 0.537)}   \\[4pt]
Respiratory failure  
                                               & 0.875 (0.867, 0.883)  & 0.571 (0.547, 0.593)  & \textbf{0.902(0.909, 0.895)}   & \textbf{0.661 (0.683, 0.638)}   \\[4pt]
Septicemia                    & 0.846 (0.838, 0.854)  & 0.507 (0.484, 0.534)  & \textbf{0.934(0.939, 0.929)}   & \textbf{0.736 (0.756, 0.715)}   \\[4pt]
Shock                                          & 0.892 (0.884, 0.901)  & 0.572 (0.546, 0.600)  & \textbf{0.932(0.938, 0.926)}   & \textbf{0.691 (0.716, 0.666)}   \\
\bottomrule
\end{tabular}}
\label{tab:pheno_subclass}
\end{table}

\begin{table}[t]
\caption{\footnotesize{\textbf{Ablation study for the model.}} The \textbf{bold} numbers represent the best AUROC in the respective column. Values are reported as mean (95\% CI).}
\centering
\resizebox{\textwidth}{!}{
\begin{tabular}{*5{c} *4{c}}
\toprule
\bfseries Setting & \bfseries Missingness & \bfseries Calibration & \bfseries Joint patching & \bfseries Unimodal & \multicolumn{2}{c}{\bfseries Mortality} & \multicolumn{2}{c}{\bfseries Clinical Conditions} \\
 & & & & & \bfseries AUROC & \bfseries AUPRC & \bfseries AUROC & \bfseries AUPRC \\
\midrule
1 & \checkmark & \checkmark & \checkmark & & 0.850 (0.834, 0.864) & 0.484 (0.446, 0.525) & 0.859 (0.850, 0.868) & 0.608 (0.585, 0.632) \\
2 & & \checkmark & \checkmark & \checkmark & 0.875 (0.861, 0.888) & 0.552 (0.514, 0.591) & 0.862 (0.853, 0.871) & 0.614 (0.591, 0.638) \\
3 & \checkmark & & \checkmark & \checkmark & 0.876 (0.863, 0.890) & 0.557 (0.518, 0.597) & 0.862 (0.853, 0.871) & 0.614 (0.591, 0.638) \\
4 & \checkmark & & & \checkmark & 0.876 (0.864, 0.888) & 0.549 (0.511, 0.586) & 0.840 (0.830, 0.849) & 0.543 (0.520, 0.567) \\
\texttt{MedPatch} & \checkmark & \checkmark & \checkmark & \checkmark & \textbf{0.876 (0.863, 0.890)} & \textbf{0.558 (0.519, 0.597)} & \textbf{0.862 (0.853, 0.871)} & \textbf{0.614 (0.591, 0.638)} \\
\bottomrule
\end{tabular}}
\label{tab:ablation}
\end{table}

\subsection{Unimodal Performance Results}
In Table \ref{tab:unimodal}, we report the performance of the unimodal models for the test sets associated with each modality. These results offer a clear demonstration of the predictive power inherent to each modality. Notably, the CXR modality exhibits the lowest predictive performance for both tasks, achieving an AUROC of 0.723 for mortality prediction and an AUROC of 0.692 for clinical conditions classification. This performance difference could be related to the smaller dataset size of CXR compared to other modalities. RR and DN demonstrate substantially improved performance, particularly for clinical conditions classification. Specifically, the DN modality achieves the highest AUROC of 0.856 and AUPRC of 0.601 — representing the best results reported to date for this task.

\subsection{Bimodal Performance Results}
We evaluated our model in the bimodal setting to facilitate comparison with other state-of-the-art baselines. The results are summarized in Table \ref{tab:bimodal}. As observed, early fusion, joint fusion, MedFuse, and MeTra exhibit comparable performance. In contrast, late fusion demonstrates noticeably inferior results across both tasks. Our proposed architecture outperforms all baselines on the mortality task, achieving an AUROC of 0.868 and an AUPRC of 0.541. It also achieves a strong performance in clinical conditions classification, achieving an AUROC of 0.773 and AUPRC of 0.439.

\subsection{Higher-Order Multimodal Performance Results}
In Table \ref{tab:multimodal}, we compare the performance of our proposed architecture against several multimodal baselines. For the mortality task, the evaluation is conducted using a trimodal configuration (i.e. EHR + CXR + RR) whereas the clinical conditions task employs all four modalities. Our results clearly indicate that \texttt{MedPatch} substantially outperforms the state-of-the-art fusion baselines, achieving an AUROC of 0.876 and AUPRC of 0.558 for in-hospital mortality prediction and an AUROC of 0.862 and AUPRC of 0.614 for clinical conditions classification. We further evaluate our architecture against four distinct ensemble baselines to emulate the multi-stage fusion. Even then, our method demonstrates significantly superior performance for both tasks. 

Additionally, we analyze the sub-class performance for the clinical conditions task in Table \ref{tab:pheno_subclass} comparing the AUROC and AUPRC achieved by \texttt{MedPatch} in the bimodal and quatrimodal settings. We note that the model performance improves across every single sub-class, with substantial improvements in AUPRC. This further reflects the predictive power associated with the DN and RR in our architecture.


\subsection{Ablations}
To evaluate the effectiveness of each of the architectural components, we run the following ablations using the trimodal mortality architecture:
\begin{enumerate}
    \item Excluding (late) unimodal predictions: We train the model and evaluate performance when fusing only the high, low, and missingness predictions without the unimodal predictions. This corresponds to $\hat{y}_{\text{late}} = \sum_{i}\tilde{\alpha}_i \hat{y}_i
$ where \(\hat{y}_i \in \{\hat{y}_{\text{high}}, \hat{y}_{\text{low}}, \hat{y}_{\text{miss}}\}\).
    \item Excluding missingness module: We train the model and evaluate performance when the late prediction does not take missingness into account, i.e, \\ 
    \(\hat{y}_i \in \{\hat{y}_{\text{high}}, \hat{y}_{\text{low}}, \hat{y}^{(ehr)}, \hat{y}^{(cxr)}, \hat{y}^{(rr)}, \hat{y}^{(dn)}\}\).
    \item Excluding calibration step: We train and evaluate the performance of the model without calibrating the confidence predictors via temperature scaling. 
    \item Excluding confidence-based patching: We train the model and evaluate performance when the late prediction does not take into account the high and low confidence outputs of the joint module, i.e, \(\hat{y}_i \in \{\hat{y}_{\text{miss}}, \hat{y}^{(ehr)}, \hat{y}^{(cxr)}, \hat{y}^{(rr)}, \hat{y}^{(dn)}\}\).
\end{enumerate}
The results are shown in Table \ref{tab:ablation}. Notably, our ablation experiments prove that the best performance is achieved when all modules are combined. The ablation studies indicate that the late fusion module contributes the most to improvements in AUROC as removing it has the biggest overall drop, bringing down the AUROC to 0.850 and the AUPRC to 0.484. This suggests the multi-stage fusion including unimodal predictions at this stage is pivotal for attaining high discriminative power. Meanwhile, the remaining modules collectively drive substantial gains in AUPRC, highlighting the effects of calibration, missingness awareness, and confidence-based token pooling on the model's ability to classify positive samples.

Further analysis in Appendix \ref{tab:weights_separate} demonstrates that the model is capable of assigning task-specific importance to different prediction components, which is indicative of its flexible and robust design. The observed variability in weighting and metric ranking highlights the significance of incorporating adaptive mechanisms that enable the model to extract nuanced insights from the confidence-separated tokens. This is a key advantage as it demonstrates a more clinically aligned multimodal learning strategy. Specifically, the adaptive weighting resembles the way clinicians alternate between individual diagnostic cues and holistic patient assessments.








\section{Discussion} 
This work introduces \texttt{MedPatch}, a multi-stage multimodal fusion framework for integrating EHR, CXR, RR, and DN, capitalizing on the unique characteristics of each modality. By employing a strategy that combines joint fusion with late fusion while explicitly handling missing data, our approach demonstrates improved performance over existing SOTA baselines in the in-hospital mortality prediction and clinical conditions classification. \texttt{MedPatch} balances efficient processing with a flexible design that mirrors the iterative and hierarchical reasoning commonly observed in clinical practice.

Our proposed approach has several strengths. First, based on the analysis of our results, we conclude that the improved performance observed with \texttt{MedPatch} stems from the architectural components that support adaptive fusion at multiple stages. The dedicated missingness module enables the model to effectively process incomplete records, an inherent challenge in clinical datasets, by providing a mechanism to weigh predictions based on modality availability. Moreover, the confidence-guided joint module and multi-stage late module complement each other and provably enhance the predictive performance of our architecture. The class-wise performance for clinical conditions and the ablation studies on missing modalities provide insight into which components of \texttt{MedPatch} drive these performance gains.

Another key strength of our study lies in the diversity of modalities used. In particular, incorporating textual data from discharge notes and radiology reports, has proven highly beneficial. As discussed in \cite{meng_data_2023} and \cite{amershi_power_2014}, the improvement achieved by including these textual modalities reinforces the view that clinical diagnosis can benefit from human-in-the-loop approaches. The use of these modalities has also enabled us to formalize and introduce new benchmark results for these clinical tasks by fusing EHR, CXR, RR, and DN data using publicly available datasets. We made our code publicly available to enable reproducibility and support future research.




\paragraph{Limitations}
Despite the promising performance of our approach, several limitations provide avenues for future investigation. Firstly, our experiments were restricted to MIMIC, which, while invaluable, may not fully capture the diversity of real-world clinical data, underscoring the need for validation across multiple and varied healthcare datasets. However, this may be challenging due to the lack of publicly available datasets that contain all modalities. Additionally, our analysis was limited to the most recent CXR collected for the patient with respect to the task prediction time, i.e. first 48 hours of admission for mortality and the full patient stay for clinical conditions classification. Future work could explore the benefit of integrating multiple CXR images collected over time for a more comprehensive patient evaluation.

We also optimized the processing pipeline for radiology reports and discharge notes for computational efficiency by employing average pooling over token patches. Although this strategy reduces computational load, it risks oversimplifying the semantic content of the full texts. Similarly, the decision to average high confidence and low confidence tokens per modality may have constrained the model’s representational power. Exploring more sophisticated patching techniques could further enhance performance.

Another important area for future work is model explainability by assessing modality-specific contributions at each stage of the architecture. Such insights could enhance clinical trust and facilitate a more targeted approach to incorporating expert knowledge during data processing. The scalable and efficient nature of \texttt{MedPatch} ensures that it can readily be extended to include additional modalities, offering a promising pathway for developing more comprehensive and interpretable multimodal deep neural networks in healthcare.




\section{Acknowledgments}
This work was supported by ASPIRE, the technology program management pillar of Abu Dhabi’s Advanced Technology Research Council (ATRC), via the ASPIRE Precision Medicine Research Institute Abu Dhabi (ASPIREPMRIAD) award grant number VRI-20-10, and the NYUAD Center for Artificial Intelligence and Robotics, funded by Tamkeen under the NYUAD Research Institute Award CG010. The research was carried out on the High Performance Computing resources at New York University Abu Dhabi. Figures \ref{Figure1} and \ref{missing} were created in BioRender (Al Jorf, B. (2025) https://BioRender.com/jl2gr7s).

\bibliography{references}

\begin{thebibliography}{34}
\providecommand{\natexlab}[1]{#1}
\providecommand{\url}[1]{\texttt{#1}}
\expandafter\ifx\csname urlstyle\endcsname\relax
  \providecommand{\doi}[1]{doi: #1}\else
  \providecommand{\doi}{doi: \begingroup \urlstyle{rm}\Url}\fi

\bibitem[Amershi et~al.(2014)Amershi, Cakmak, Knox, and Kulesza]{amershi_power_2014}
Saleema Amershi, Maya Cakmak, William~Bradley Knox, and Todd Kulesza.
\newblock Power to the {People}: {The} {Role} of {Humans} in {Interactive} {Machine} {Learning}.
\newblock \emph{AI Magazine}, 35\penalty0 (4):\penalty0 105--120, December 2014.
\newblock ISSN 2371-9621.
\newblock \doi{10.1609/aimag.v35i4.2513}.
\newblock URL \url{https://ojs.aaai.org/aimagazine/index.php/aimagazine/article/view/2513}.
\newblock Number: 4.

\bibitem[Baliah et~al.(2023)Baliah, Maani, Sanjeev, and Khan]{baliah_exploring_2023}
Sanoojan Baliah, Fadillah~A. Maani, Santosh Sanjeev, and Muhammad~Haris Khan.
\newblock Exploring the {Transfer} {Learning} {Capabilities} of {CLIP} in {Domain} {Generalization} for {Diabetic} {Retinopathy}.
\newblock In \emph{Machine {Learning} in {Medical} {Imaging}: 14th {International} {Workshop}, {MLMI} 2023, {Held} in {Conjunction} with {MICCAI} 2023, {Vancouver}, {BC}, {Canada}, {October} 8, 2023, {Proceedings}, {Part} {I}}, pages 444--453, Berlin, Heidelberg, October 2023. Springer-Verlag.
\newblock ISBN 978-3-031-45672-5.
\newblock \doi{10.1007/978-3-031-45673-2_44}.
\newblock URL \url{https://doi.org/10.1007/978-3-031-45673-2_44}.

\bibitem[Demner-Fushman et~al.(2009)Demner-Fushman, Chapman, and McDonald]{demner-fushman_what_2009}
Dina Demner-Fushman, Wendy~W. Chapman, and Clement~J. McDonald.
\newblock What can {Natural} {Language} {Processing} do for {Clinical} {Decision} {Support}?
\newblock \emph{Journal of biomedical informatics}, 42\penalty0 (5):\penalty0 760--772, October 2009.
\newblock ISSN 1532-0464.
\newblock \doi{10.1016/j.jbi.2009.08.007}.
\newblock URL \url{https://www.ncbi.nlm.nih.gov/pmc/articles/PMC2757540/}.

\bibitem[Deznabi et~al.(2021)Deznabi, Iyyer, and Fiterau]{deznabi_predicting_2021}
Iman Deznabi, Mohit Iyyer, and Madalina Fiterau.
\newblock Predicting in-hospital mortality by combining clinical notes with time-series data.
\newblock In Chengqing Zong, Fei Xia, Wenjie Li, and Roberto Navigli, editors, \emph{Findings of the {Association} for {Computational} {Linguistics}: {ACL}-{IJCNLP} 2021}, pages 4026--4031, Online, August 2021. Association for Computational Linguistics.
\newblock \doi{10.18653/v1/2021.findings-acl.352}.
\newblock URL \url{https://aclanthology.org/2021.findings-acl.352}.

\bibitem[Eslami et~al.(2023)Eslami, Meinel, and de~Melo]{eslami_pubmedclip_2023}
Sedigheh Eslami, Christoph Meinel, and Gerard de~Melo.
\newblock {PubMedCLIP}: {How} {Much} {Does} {CLIP} {Benefit} {Visual} {Question} {Answering} in the {Medical} {Domain}?
\newblock In Andreas Vlachos and Isabelle Augenstein, editors, \emph{Findings of the {Association} for {Computational} {Linguistics}: {EACL} 2023}, pages 1181--1193, Dubrovnik, Croatia, May 2023. Association for Computational Linguistics.
\newblock \doi{10.18653/v1/2023.findings-eacl.88}.
\newblock URL \url{https://aclanthology.org/2023.findings-eacl.88/}.

\bibitem[Hayat et~al.(2022)Hayat, Geras, and Shamout]{hayat_medfuse_2022}
Nasir Hayat, Krzysztof~J. Geras, and Farah~E. Shamout.
\newblock {MedFuse}: {Multi}-modal fusion with clinical time-series data and chest {X}-ray images.
\newblock In \emph{Proceedings of the 7th {Machine} {Learning} for {Healthcare} {Conference}}, pages 479--503. PMLR, December 2022.
\newblock URL \url{https://proceedings.mlr.press/v182/hayat22a.html}.
\newblock ISSN: 2640-3498.

\bibitem[He et~al.(2025)He, Huang, Jiang, Nie, Wang, Wang, and Chen]{he_foundation_2025}
Yuting He, Fuxiang Huang, Xinrui Jiang, Yuxiang Nie, Minghao Wang, Jiguang Wang, and Hao Chen.
\newblock Foundation {Model} for {Advancing} {Healthcare}: {Challenges}, {Opportunities} and {Future} {Directions}.
\newblock \emph{IEEE Reviews in Biomedical Engineering}, 18:\penalty0 172--191, 2025.
\newblock ISSN 1941-1189.
\newblock \doi{10.1109/RBME.2024.3496744}.
\newblock URL \url{https://ieeexplore.ieee.org/document/10750441/}.

\bibitem[Hemker et~al.(2024)Hemker, Simidjievski, and Jamnik]{hemker_healnet_2024}
Konstantin Hemker, Nikola Simidjievski, and Mateja Jamnik.
\newblock {HEALNet}: {Multimodal} {Fusion} for {Heterogeneous} {Biomedical} {Data}.
\newblock November 2024.
\newblock URL \url{https://openreview.net/forum?id=HUxtJcQpDS}.

\bibitem[Huang et~al.(2020)Huang, Pareek, Seyyedi, Banerjee, and Lungren]{huang_fusion_2020}
Shih-Cheng Huang, Anuj Pareek, Saeed Seyyedi, Imon Banerjee, and Matthew~P. Lungren.
\newblock Fusion of medical imaging and electronic health records using deep learning: a systematic review and implementation guidelines.
\newblock \emph{npj Digital Medicine}, 3\penalty0 (1):\penalty0 1--9, October 2020.
\newblock ISSN 2398-6352.
\newblock \doi{10.1038/s41746-020-00341-z}.
\newblock URL \url{https://www.nature.com/articles/s41746-020-00341-z}.

\bibitem[Huang et~al.(2024)Huang, Jensen, Yeung-Levy, Lungren, Poon, and Chaudhari]{huang_multimodal_2024}
Shih-Cheng Huang, Malte Jensen, Serena Yeung-Levy, Matthew~P. Lungren, Hoifung Poon, and Akshay~S. Chaudhari.
\newblock Multimodal {Foundation} {Models} for {Medical} {Imaging} - {A} {Systematic} {Review} and {Implementation} {Guidelines}, October 2024.
\newblock URL \url{https://www.medrxiv.org/content/10.1101/2024.10.23.24316003v1}.
\newblock Pages: 2024.10.23.24316003.

\bibitem[Johnson et~al.(2023{\natexlab{a}})Johnson, Pollard, Horng, Celi, and Mark]{johnson_mimic-iv-note_2023}
Alistair Johnson, Tom Pollard, Steven Horng, Leo~Anthony Celi, and Roger Mark.
\newblock {MIMIC}-{IV}-{Note}: {Deidentified} free-text clinical notes, 2023{\natexlab{a}}.
\newblock URL \url{https://physionet.org/content/mimic-iv-note/2.2/}.

\bibitem[Johnson et~al.(2019)Johnson, Pollard, Berkowitz, Greenbaum, Lungren, Deng, Mark, and Horng]{johnson_mimic-cxr_2019}
Alistair E.~W. Johnson, Tom~J. Pollard, Seth~J. Berkowitz, Nathaniel~R. Greenbaum, Matthew~P. Lungren, Chih-ying Deng, Roger~G. Mark, and Steven Horng.
\newblock {MIMIC}-{CXR}, a de-identified publicly available database of chest radiographs with free-text reports.
\newblock \emph{Scientific Data}, 6\penalty0 (1):\penalty0 317, December 2019.
\newblock ISSN 2052-4463.
\newblock \doi{10.1038/s41597-019-0322-0}.
\newblock URL \url{https://www.nature.com/articles/s41597-019-0322-0}.
\newblock Publisher: Nature Publishing Group.

\bibitem[Johnson et~al.(2023{\natexlab{b}})Johnson, Bulgarelli, Shen, Gayles, Shammout, Horng, Pollard, Hao, Moody, Gow, Lehman, Celi, and Mark]{johnson_mimic-iv_2023}
Alistair E.~W. Johnson, Lucas Bulgarelli, Lu~Shen, Alvin Gayles, Ayad Shammout, Steven Horng, Tom~J. Pollard, Sicheng Hao, Benjamin Moody, Brian Gow, Li-wei~H. Lehman, Leo~A. Celi, and Roger~G. Mark.
\newblock {MIMIC}-{IV}, a freely accessible electronic health record dataset.
\newblock \emph{Scientific Data}, 10\penalty0 (1):\penalty0 1, January 2023{\natexlab{b}}.
\newblock ISSN 2052-4463.
\newblock \doi{10.1038/s41597-022-01899-x}.
\newblock URL \url{https://www.nature.com/articles/s41597-022-01899-x}.

\bibitem[Khader et~al.(2023)Khader, Kather, Müller-Franzes, Wang, Han, Tayebi~Arasteh, Hamesch, Bressem, Haarburger, Stegmaier, Kuhl, Nebelung, and Truhn]{khader_medical_2023}
Firas Khader, Jakob~Nikolas Kather, Gustav Müller-Franzes, Tianci Wang, Tianyu Han, Soroosh Tayebi~Arasteh, Karim Hamesch, Keno Bressem, Christoph Haarburger, Johannes Stegmaier, Christiane Kuhl, Sven Nebelung, and Daniel Truhn.
\newblock Medical transformer for multimodal survival prediction in intensive care: integration of imaging and non-imaging data.
\newblock \emph{Scientific Reports}, 13\penalty0 (1):\penalty0 10666, July 2023.
\newblock ISSN 2045-2322.
\newblock \doi{10.1038/s41598-023-37835-1}.
\newblock URL \url{https://www.nature.com/articles/s41598-023-37835-1}.

\bibitem[Krones et~al.(2025)Krones, Marikkar, Parsons, Szmul, and Mahdi]{krones_review_2025}
Felix Krones, Umar Marikkar, Guy Parsons, Adam Szmul, and Adam Mahdi.
\newblock Review of multimodal machine learning approaches in healthcare.
\newblock \emph{Information Fusion}, 114:\penalty0 102690, February 2025.
\newblock ISSN 1566-2535.
\newblock \doi{10.1016/j.inffus.2024.102690}.
\newblock URL \url{https://www.sciencedirect.com/science/article/pii/S1566253524004688}.

\bibitem[Lee et~al.(2020)Lee, Yoon, Kim, Kim, Kim, So, and Kang]{lee_biobert_2020}
Jinhyuk Lee, Wonjin Yoon, Sungdong Kim, Donghyeon Kim, Sunkyu Kim, Chan~Ho So, and Jaewoo Kang.
\newblock {BioBERT}: a pre-trained biomedical language representation model for biomedical text mining.
\newblock \emph{Bioinformatics}, 36\penalty0 (4):\penalty0 1234--1240, February 2020.
\newblock ISSN 1367-4803.
\newblock \doi{10.1093/bioinformatics/btz682}.
\newblock URL \url{https://doi.org/10.1093/bioinformatics/btz682}.

\bibitem[Lee et~al.(2023)Lee, Lee, Hahn, Hyun, Choi, Ahn, and Lee]{lee_learning_2023}
Kwanhyung Lee, Soojeong Lee, Sangchul Hahn, Heejung Hyun, Edward Choi, Byungeun Ahn, and Joohyung Lee.
\newblock Learning {Missing} {Modal} {Electronic} {Health} {Records} with {Unified} {Multi}-modal {Data} {Embedding} and {Modality}-{Aware} {Attention}.
\newblock In \emph{Proceedings of the 8th {Machine} {Learning} for {Healthcare} {Conference}}, pages 423--442. PMLR, December 2023.
\newblock URL \url{https://proceedings.mlr.press/v219/lee23a.html}.
\newblock ISSN: 2640-3498.

\bibitem[Lin et~al.(2021)Lin, Wang, Ding, Zhao, Wang, and Peng]{lin_empirical_2021}
Mingquan Lin, Song Wang, Ying Ding, Lihui Zhao, Fei Wang, and Yifan Peng.
\newblock An empirical study of using radiology reports and images to improve {ICU}-mortality prediction.
\newblock \emph{Proceedings. IEEE International Conference on Healthcare Informatics}, 2021:\penalty0 497--498, August 2021.
\newblock ISSN 2575-2626.
\newblock \doi{10.1109/ichi52183.2021.00088}.
\newblock URL \url{https://www.ncbi.nlm.nih.gov/pmc/articles/PMC9076267/}.

\bibitem[Liu et~al.(2025)Liu, Cheng, Shi, Shah, Bai, and Arcucci]{liu_imitate_2025}
Che Liu, Sibo Cheng, Miaojing Shi, Anand Shah, Wenjia Bai, and Rossella Arcucci.
\newblock {IMITATE}: {Clinical} {Prior} {Guided} {Hierarchical} {Vision}-{Language} {Pre}-{Training}.
\newblock \emph{IEEE Transactions on Medical Imaging}, 44\penalty0 (1):\penalty0 519--529, January 2025.
\newblock ISSN 1558-254X.
\newblock \doi{10.1109/TMI.2024.3449690}.
\newblock URL \url{https://ieeexplore.ieee.org/document/10646593}.

\bibitem[Liu et~al.(2023)Liu, Capurro, Nguyen, and Verspoor]{liu_attention-based_2023}
Jinghui Liu, Daniel Capurro, Anthony Nguyen, and Karin Verspoor.
\newblock Attention-based multimodal fusion with contrast for robust clinical prediction in the face of missing modalities.
\newblock \emph{Journal of Biomedical Informatics}, 145:\penalty0 104466, September 2023.
\newblock ISSN 1532-0464.
\newblock \doi{10.1016/j.jbi.2023.104466}.
\newblock URL \url{https://www.sciencedirect.com/science/article/pii/S1532046423001879}.

\bibitem[Lyu et~al.(2023)Lyu, Dong, Wong, Zheng, Abell-Hart, Wang, and Chen]{lyu_multimodal_2023}
Weimin Lyu, Xinyu Dong, Rachel Wong, Songzhu Zheng, Kayley Abell-Hart, Fusheng Wang, and Chao Chen.
\newblock A {Multimodal} {Transformer}: {Fusing} {Clinical} {Notes} with {Structured} {EHR} {Data} for {Interpretable} {In}-{Hospital} {Mortality} {Prediction}.
\newblock \emph{AMIA Annual Symposium Proceedings}, 2022:\penalty0 719--728, April 2023.
\newblock ISSN 1942-597X.
\newblock URL \url{https://www.ncbi.nlm.nih.gov/pmc/articles/PMC10148371/}.

\bibitem[Meng(2023)]{meng_data_2023}
Xiao-Li Meng.
\newblock Data {Science} and {Engineering} {With} {Human} in the {Loop}, {Behind} the {Loop}, and {Above} the {Loop}.
\newblock \emph{Harvard Data Science Review}, 5\penalty0 (2), April 2023.
\newblock ISSN 2644-2353, 2688-8513.
\newblock \doi{10.1162/99608f92.68a012eb}.
\newblock URL \url{https://hdsr.mitpress.mit.edu/pub/812vijgg/release/3}.
\newblock Publisher: The MIT Press.

\bibitem[Munro and Swamy(2024)]{munro_documentation_2024}
Cindy~L. Munro and Lakshman Swamy.
\newblock Documentation, {Data}, and {Decision}-{Making}.
\newblock \emph{American Journal of Critical Care}, 33\penalty0 (3):\penalty0 162--165, May 2024.
\newblock ISSN 1062-3264.
\newblock \doi{10.4037/ajcc2024617}.
\newblock URL \url{https://aacnjournals.org/ajcconline/article/33/3/162/32425/Documentation-Data-and-Decision-Making}.
\newblock Publisher: American Association of Critical-Care Nurses.

\bibitem[Nawrot et~al.(2023)Nawrot, Chorowski, Lancucki, and Ponti]{nawrot_efficient_2023}
Piotr Nawrot, Jan Chorowski, Adrian Lancucki, and Edoardo~Maria Ponti.
\newblock Efficient {Transformers} with {Dynamic} {Token} {Pooling}.
\newblock In Anna Rogers, Jordan Boyd-Graber, and Naoaki Okazaki, editors, \emph{Proceedings of the 61st {Annual} {Meeting} of the {Association} for {Computational} {Linguistics} ({Volume} 1: {Long} {Papers})}, pages 6403--6417, Toronto, Canada, July 2023. Association for Computational Linguistics.
\newblock \doi{10.18653/v1/2023.acl-long.353}.
\newblock URL \url{https://aclanthology.org/2023.acl-long.353/}.

\bibitem[Niu et~al.(2023)Niu, Zhang, Peng, Pan, and Xiao]{niu_deep_2023}
Ke~Niu, Ke~Zhang, Xueping Peng, Yijie Pan, and Naian Xiao.
\newblock Deep multi-modal intermediate fusion of clinical record and time series data in mortality prediction.
\newblock \emph{Frontiers in Molecular Biosciences}, 10, March 2023.
\newblock ISSN 2296-889X.
\newblock \doi{10.3389/fmolb.2023.1136071}.
\newblock URL \url{https://www.frontiersin.org/articles/10.3389/fmolb.2023.1136071}.

\bibitem[Pagnoni et~al.(2024)Pagnoni, Pasunuru, Rodriguez, Nguyen, Muller, Li, Zhou, Yu, Weston, Zettlemoyer, Ghosh, Lewis, Holtzman, and Iyer]{pagnoni_byte_2024}
Artidoro Pagnoni, Ram Pasunuru, Pedro Rodriguez, John Nguyen, Benjamin Muller, Margaret Li, Chunting Zhou, Lili Yu, Jason Weston, Luke Zettlemoyer, Gargi Ghosh, Mike Lewis, Ari Holtzman, and Srinivasan Iyer.
\newblock Byte {Latent} {Transformer}: {Patches} {Scale} {Better} {Than} {Tokens}.
\newblock 2024.

\bibitem[Pham et~al.(2024)Pham, Brecheisen, Nguyen, Nguyen, and Le]{pham_i-ai_2024}
Trong~Thang Pham, Jacob Brecheisen, Anh Nguyen, Hien Nguyen, and Ngan Le.
\newblock I-{AI}: {A} {Controllable} \& {Interpretable} {AI} {System} for {Decoding} {Radiologists}’ {Intense} {Focus} for {Accurate} {CXR} {Diagnoses}.
\newblock In \emph{2024 {IEEE}/{CVF} {Winter} {Conference} on {Applications} of {Computer} {Vision} ({WACV})}, pages 7835--7844, January 2024.
\newblock \doi{10.1109/WACV57701.2024.00767}.
\newblock URL \url{https://ieeexplore.ieee.org/document/10483737/}.
\newblock ISSN: 2642-9381.

\bibitem[Shamout et~al.(2021)Shamout, Shen, Wu, Kaku, Park, Makino, Jastrzębski, Witowski, Wang, Zhang, Dogra, Cao, Razavian, Kudlowitz, Azour, Moore, Lui, Aphinyanaphongs, Fernandez-Granda, and Geras]{shamout_artificial_2021}
Farah~E. Shamout, Yiqiu Shen, Nan Wu, Aakash Kaku, Jungkyu Park, Taro Makino, Stanisław Jastrzębski, Jan Witowski, Duo Wang, Ben Zhang, Siddhant Dogra, Meng Cao, Narges Razavian, David Kudlowitz, Lea Azour, William Moore, Yvonne~W. Lui, Yindalon Aphinyanaphongs, Carlos Fernandez-Granda, and Krzysztof~J. Geras.
\newblock An artificial intelligence system for predicting the deterioration of {COVID}-19 patients in the emergency department.
\newblock \emph{npj Digital Medicine}, 4\penalty0 (1):\penalty0 1--11, May 2021.
\newblock ISSN 2398-6352.
\newblock \doi{10.1038/s41746-021-00453-0}.
\newblock URL \url{https://www.nature.com/articles/s41746-021-00453-0}.
\newblock Publisher: Nature Publishing Group.

\bibitem[Steinberg et~al.(2021)Steinberg, Jung, Fries, Corbin, Pfohl, and Shah]{steinberg_language_2021}
Ethan Steinberg, Ken Jung, Jason~A. Fries, Conor~K. Corbin, Stephen~R. Pfohl, and Nigam~H. Shah.
\newblock Language models are an effective representation learning technique for electronic health record data.
\newblock \emph{Journal of Biomedical Informatics}, 113:\penalty0 103637, January 2021.
\newblock ISSN 1532-0464.
\newblock \doi{10.1016/j.jbi.2020.103637}.
\newblock URL \url{https://www.sciencedirect.com/science/article/pii/S1532046420302653}.

\bibitem[Wan et~al.(2023)Wan, Liu, Zhang, Fu, Wang, Cheng, Ma, Quilodrán-Casas, and Arcucci]{wan_med-unic_2023}
Zhongwei Wan, Che Liu, Mi~Zhang, Jie Fu, Benyou Wang, Sibo Cheng, Lei Ma, César Quilodrán-Casas, and Rossella Arcucci.
\newblock Med-{UniC}: {Unifying} {Cross}-{Lingual} {Medical} {Vision}-{Language} {Pre}-{Training} by {Diminishing} {Bias}.
\newblock \emph{Advances in Neural Information Processing Systems}, 36:\penalty0 56186--56197, December 2023.
\newblock URL \url{https://proceedings.neurips.cc/paper_files/paper/2023/hash/af38fb8e90d586f209235c94119ba193-Abstract-Conference.html}.

\bibitem[Wu et~al.(2023)Wu, Zhang, Zhang, Wang, and Xie]{wu_medklip_2023}
Chaoyi Wu, Xiaoman Zhang, Ya~Zhang, Yanfeng Wang, and Weidi Xie.
\newblock {MedKLIP}: {Medical} {Knowledge} {Enhanced} {Language}-{Image} {Pre}-{Training} for {X}-ray {Diagnosis}.
\newblock In \emph{2023 {IEEE}/{CVF} {International} {Conference} on {Computer} {Vision} ({ICCV})}, pages 21315--21326, Paris, France, October 2023. IEEE.
\newblock ISBN 9798350307184.
\newblock \doi{10.1109/ICCV51070.2023.01954}.
\newblock URL \url{https://ieeexplore.ieee.org/document/10376864/}.

\bibitem[Xu et~al.(2021)Xu, So, and Dai]{xu_mufasa_2021}
Zhen Xu, David~R. So, and Andrew~M. Dai.
\newblock {MUFASA}: {Multimodal} {Fusion} {Architecture} {Search} for {Electronic} {Health} {Records}.
\newblock \emph{Proceedings of the AAAI Conference on Artificial Intelligence}, 35\penalty0 (12):\penalty0 10532--10540, May 2021.
\newblock ISSN 2374-3468.
\newblock \doi{10.1609/aaai.v35i12.17260}.
\newblock URL \url{https://ojs.aaai.org/index.php/AAAI/article/view/17260}.
\newblock Number: 12.

\bibitem[Yang et~al.(2021)Yang, Kuang, and Xia]{yang_multimodal_2021}
Haiyang Yang, Li~Kuang, and FengQiang Xia.
\newblock Multimodal temporal-clinical note network for mortality prediction.
\newblock \emph{Journal of Biomedical Semantics}, 12\penalty0 (1):\penalty0 3, February 2021.
\newblock ISSN 2041-1480.
\newblock \doi{10.1186/s13326-021-00235-3}.
\newblock URL \url{https://doi.org/10.1186/s13326-021-00235-3}.

\bibitem[Yao et~al.(2024)Yao, Yin, Cheung, Liu, and Qin]{yao_drfuse_2024}
Wenfang Yao, Kejing Yin, William~K. Cheung, Jia Liu, and Jing Qin.
\newblock {DrFuse}: {Learning} {Disentangled} {Representation} for {Clinical} {Multi}-{Modal} {Fusion} with {Missing} {Modality} and {Modal} {Inconsistency}.
\newblock \emph{Proceedings of the AAAI Conference on Artificial Intelligence}, 38\penalty0 (15):\penalty0 16416--16424, March 2024.
\newblock ISSN 2374-3468.
\newblock \doi{10.1609/aaai.v38i15.29578}.
\newblock URL \url{https://ojs.aaai.org/index.php/AAAI/article/view/29578}.
\newblock Number: 15.

\end{thebibliography}

\newpage
\appendix

\section{Missingness Module}

We provide a more detailed look at \texttt{MedPatch}'s missingness module in Figure \ref{missing}. The predictions from this module are sent to the late module for each patient, where they are used in the final weighted average output. 
\renewcommand{\thefigure}{\Alph{section}\arabic{figure}}
\setcounter{figure}{0}
\renewcommand{\thetable}{\Alph{section}\arabic{table}}
\setcounter{table}{0}
\begin{figure}[htbp]
    \centering
    \includegraphics[width=1.0\textwidth]{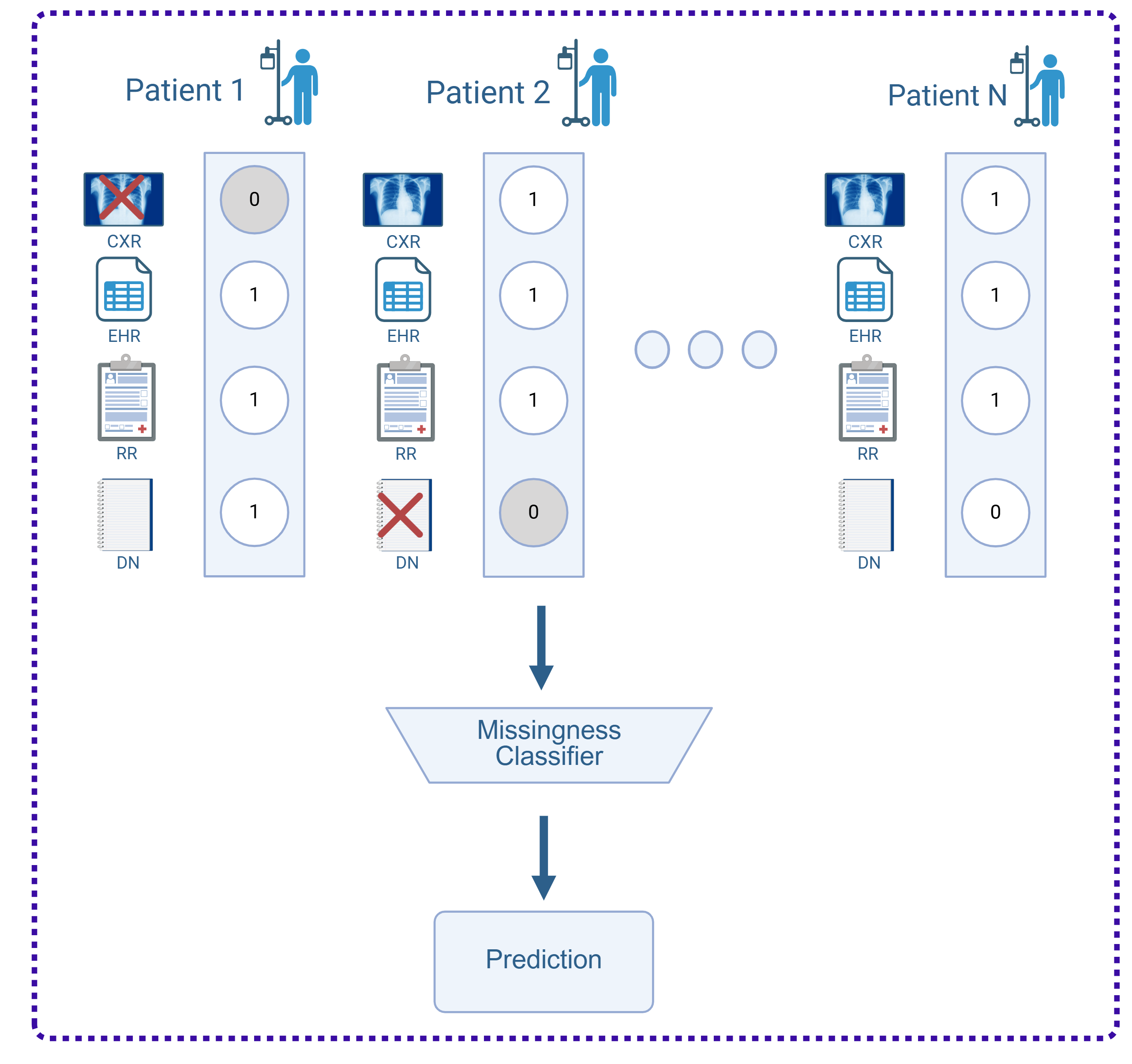} 
    \caption{Overview of the missingness module. Each patient is assigned a missingness vector which indicates the available modalities. These vectors are then passed onto a missingness classifier that learns to associate missingness patterns with specific labels.}\vspace{-5mm}
    \label{missing}
\end{figure}

\newpage

\section{Dataset Summary}
Table \ref{EHR} summarizes the measurements used in each patients electronic health records and Table \ref{Prev} contains the positive rates for each patient condition across our data splits. In Table \ref{age}, we summarize the age groups of the patients in the dataset. Table \ref{Gender} shows the gender distribution between Male (M) and Female (F). Figures \ref{fig:mortality_diff} and \ref{fig:pheno_dff} show the time trends in CXR collection for the different tasks.
\renewcommand{\thefigure}{\Alph{section}\arabic{figure}}
\setcounter{figure}{0}
\renewcommand{\thetable}{\Alph{section}\arabic{table}}
\setcounter{table}{0}

\begin{table}[ht]
\centering
\small
\caption{Variable Names, Descriptions, Source Tables, and Impute Values}
\resizebox{\textwidth}{!}{
\begin{tabular}{cllll}
\toprule
\# & \textbf{Variable Name} & \textbf{Variable Description} & \textbf{Source Table} & \textbf{Impute Value} \\
\midrule
\multicolumn{5}{l}{\textbf{Categorical Variables}} \\
1 & Capillary Refill Rate & Indicator of circulatory system function & chartevents & 0.0 \\
2 & Glasgow Coma Scale - Eye Opening & Assesses eye response to stimuli & chartevents & 4 Spontaneously \\
3 & Glasgow Coma Scale - Motor Response & Assesses motor response to stimuli & chartevents & 6 Obeys Commands \\
4 & Glasgow Coma Scale - Verbal Response & Assesses verbal response to stimuli & chartevents & 5 Oriented \\
5 & Glasgow Coma Scale - Total & Overall assessment of consciousness level & chartevents & 15 \\
\midrule
\multicolumn{5}{l}{\textbf{Continuous Variables}} \\
6 & Diastolic Blood Pressure & Blood pressure during heart’s relaxation phase & chartevents & 59.0 \\
7 & Fraction of Inspired Oxygen & Oxygen concentration in inhaled air & chartevents & 0.21 \\
8 & Glucose & Blood sugar level & labevents & 128.0 \\
9 & Heart Rate & Number of heartbeats per minute & chartevents & 86 \\
10 & Height & Patient’s height & chartevents & 170.0 \\
11 & Mean Blood Pressure & Average blood pressure during a single cardiac cycle & chartevents & 77.0 \\
12 & Oxygen Saturation & Percentage of oxygen-saturated hemoglobin & chartevents & 98.0 \\
13 & Respiratory Rate & Number of breaths per minute & chartevents & 19 \\
14 & Systolic Blood Pressure & Blood pressure during heart’s contraction phase & chartevents & 118.0 \\
15 & Temperature & Body temperature & chartevents & 36.6 \\
16 & Weight & Patient’s weight & chartevents & 81.0 \\
17 & pH & Acidity or alkalinity of the blood & labevents & 7.4 \\
\bottomrule
\end{tabular}
}
\label{EHR}
\end{table}

\begin{table}[h]
\centering
\caption{Prevalence of clinical conditions and in-hospital mortality across training, validation, and test datasets.}
\resizebox{0.69\textwidth}{!}{
\begin{tabular}{@{}lccc@{}}
\toprule
\textbf{Condition} & \textbf{Train \%} & \textbf{Val \%} & \textbf{Test \%} \\
\midrule
\multicolumn{4}{l}{\textbf{In-hospital Mortality}} \\
\midrule
Patient Mortality Rate & 12.5 & 11.9 & 12.4 \\
\midrule
\multicolumn{4}{l}{\textbf{Clinical Conditions}} \\
\midrule
Acute and unspecified renal failure & 26.9 & 26.1 & 26.9 \\
Acute cerebrovascular disease & 5.6 & 5.0 & 5.7 \\
Acute myocardial infarction & 7.5 & 7.6 & 7.7 \\
Cardiac dysrhythmias & 32.6 & 31.1 & 32.4 \\
Chronic kidney disease & 20.6 & 20.8 & 21.0 \\
Chronic obstructive pulmonary disease and bronchiectasis & 14.3 & 14.6 & 14.1 \\
Complications of surgical procedures or medical care & 18.9 & 19.4 & 18.3 \\
Conduction disorders & 10.0 & 10.6 & 10.2 \\
Congestive heart failure; nonhypertensive & 25.5 & 25.7 & 25.0 \\
Coronary atherosclerosis and other heart disease & 31.1 & 31.8 & 31.7 \\
Diabetes mellitus with complications & 11.4 & 12.2 & 11.1 \\
Diabetes mellitus without complication & 17.2 & 17.3 & 17.3 \\
Disorders of lipid metabolism & 40.5 & 41.7 & 40.6 \\
Essential hypertension & 41.8 & 41.3 & 42.0 \\
Fluid and electrolyte disorders & 37.2 & 37.0 & 37.2 \\
Gastrointestinal hemorrhage & 7.0 & 6.7 & 7.2 \\
Hypertension with complications and secondary hypertension & 21.5 & 21.9 & 21.7 \\
Other liver diseases & 12.5 & 12.4 & 12.5 \\
Other lower respiratory disease & 9.5 & 9.8 & 9.5 \\
Other upper respiratory disease & 4.8 & 5.7 & 4.7 \\
Pleurisy; pneumothorax; pulmonary collapse & 6.7 & 6.7 & 6.9 \\
Pneumonia (except that caused by tuberculosis or sexually transmitted disease) & 12.7 & 12.4 & 12.3 \\
Respiratory failure; insufficiency; arrest (adult) & 16.0 & 16.7 & 15.6 \\
Septicemia (except in labor) & 15.8 & 15.4 & 15.6 \\
Shock & 12.3 & 12.3 & 12.0 \\
\bottomrule
\end{tabular}
}
\label{Prev}
\end{table}

\newpage

\begin{minipage}{.48\linewidth}
\centering
\captionof{table}{Age distribution across the MIMIC dataset.}\label{tab:mimic-age}
\vspace{2mm}
\begin{tabular}{@{}l r@{}}
\toprule
\multicolumn{2}{@{}l}{\textbf{Age bucket counts:}}\\[4pt]
\midrule
0--20 & 561\\
21--40 & 6,096\\
41--60 & 17,336\\
61--80 & 25,385\\
80+    & 9,994\\
\bottomrule
\label{age}
\end{tabular}
\end{minipage}\hfill
\begin{minipage}{.48\linewidth}
\centering
\captionof{table}{Gender distribution across the MIMIC dataset.}\label{tab:mimic-gender}
\vspace{2mm}
\begin{tabular}{@{}l r@{}}
\toprule
\multicolumn{2}{@{}l}{\textbf{Gender distribution:}}\\[4pt]
\midrule
M & 33,113\\
F & 26,259\\
\bottomrule
\label{Gender}
\end{tabular}
\end{minipage}

\begin{figure}
\centering
\includegraphics[width=0.5\linewidth]{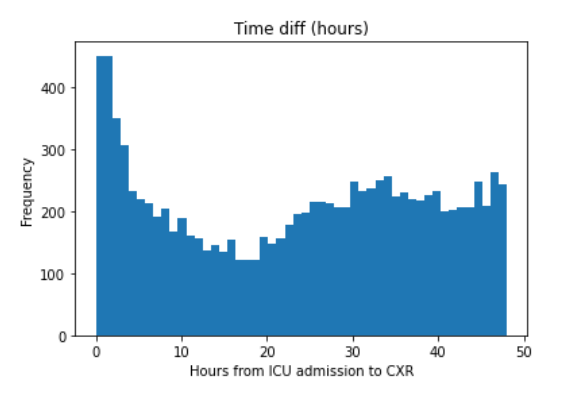}
\caption{CXR collection time since admission for the in-hospital mortality prediction task. CXRs after 48 hours are not accepted.}
\label{fig:mortality_diff}
\end{figure}\hfill

\begin{figure}
\centering
\includegraphics[width=0.5\linewidth]{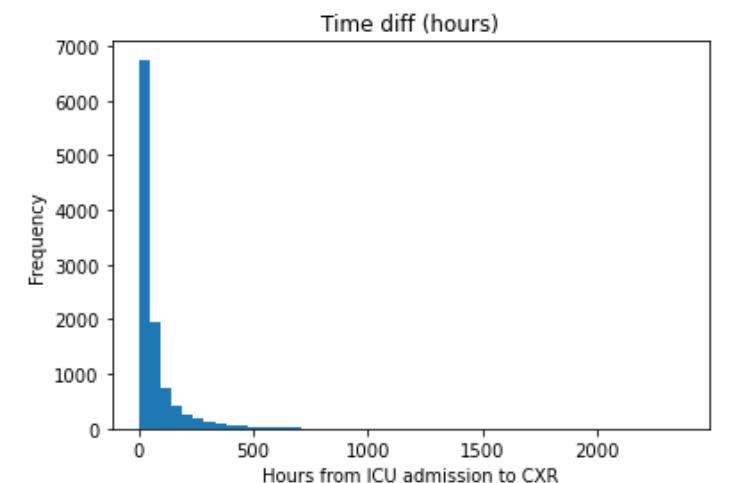}
\caption{CXR collection time since admission for the clinical condition classification task.}
\label{fig:pheno_dff}
\end{figure}

\newpage

\section{Additional Results}
\renewcommand{\thetable}{\Alph{section}\arabic{table}}
\setcounter{table}{0}

\setcounter{figure}{0}
\renewcommand{\thefigure}{\Alph{section}\arabic{figure}}

\subsection{Detailed MedPatch Metrics}
To effectively evaluate our model, we look at various combinations of the late, high, and low predictions from our model, and use them to calculate three different sets of metrics:

\begin{itemize}
    \item Late: AUROC and AUPRC computed from the late prediction \(\hat{y}_{\text{late}}\).
    \item Combined: AUROC and AUPRC computed using the average prediction of high-confidence, late, and low-confidence predictions:

    \[
    \hat{y}_{\text{combined}} = \frac{\hat{y}_{\text{high}} + \hat{y}_{\text{late}} + \hat{y}_{\text{low}}}{3}.
    \]
    \item Reduced Combined: AUROC and AUPRC computed using only the average of high-confidence and late predictions, explicitly excluding low-confidence predictions:
    \[
    \hat{y}_{\text{reduced}} = \frac{\hat{y}_{\text{high}} + \hat{y}_{\text{late}}}{2}.
    \]
\end{itemize}

Table~\ref{tab:auroc_auprc} summarizes the AUROC and AUPRC values for both the bimodal and multimodal settings across the two clinical tasks (In-hospital Mortality and Phenotyping). The bold numbers indicate the best performance per metric within each row, demonstrating that generally the late prediction generally achieves the highest scores, but for the quatrimodal phenotyping task the combined performs best.

\begin{table}[h]
\caption{Comparison of AUROC and AUPRC metrics in \texttt{MedPatch}'s bimodal and multimodal settings, split by In-hospital Mortality and Phenotyping tasks. The bold results are the highest per metric group (AUROC, AUPRC) per row.}
\begin{center}
\resizebox{0.9\linewidth}{!}{%
\begin{tabular}{l ccc ccc}
\toprule
{\textbf{Setting}} 
    & \multicolumn{3}{c}{\textbf{AUROC}} 
    & \multicolumn{3}{c}{\textbf{AUPRC}} \\
\cmidrule(lr){2-4} \cmidrule(lr){5-7}
    & \textbf{Late} & \textbf{Combined} & \textbf{Reduced} 
    & \textbf{Late} & \textbf{Combined} & \textbf{Reduced} \\
\midrule
\multicolumn{7}{c}{\textbf{In-hospital Mortality}}\\
\midrule
Bimodal    & \textbf{0.868} & 0.850  & 0.861 & \textbf{0.541} & 0.510 & 0.521 \\
Multimodal & \textbf{0.876} & 0.867  & 0.861 & \textbf{0.558} & 0.519 & 0.523 \\
\midrule
\multicolumn{7}{c}{\textbf{Phenotyping}}\\
\midrule
Bimodal    & \textbf{0.773} & 0.701 & 0.699 & \textbf{0.439} & 0.328 & 0.323 \\
Multimodal & 0.818 & \textbf{0.862} & 0.856 & 0.518 & \textbf{0.614} & 0.602 \\
\bottomrule
\end{tabular}}
\end{center}
\label{tab:auroc_auprc}
\end{table}

\newpage
\subsection{Clinical Conditions Alpha Weights Comparison}

An important component of the \texttt{MedPatch} framework is the adaptive weighting of modality-specific and confidence-based predictors within the late fusion module. In this section, we analyze the learned alpha weights assigned to each prediction component for clinical condition prediction.

Table~\ref{tab:pheno_alpha} (reproduced for alpha weights) shows the alpha weights per clinical condition class. Each row corresponds to a diagnostic category, with weights assigned to unimodal predictors (EHR, CXR, DN, RR), the missingness indicator, and the confidence-derived low and high predictions. These alpha weights reflect the relative importance that the model assigns to each modality and the confidence levels for predicting a given clinical condition. For instance, a higher alpha weight for a modality such as DN in one condition may indicate that the textual information in the discharge note is particularly predictive for that condition.

\begin{table}[h]
\caption{Comparison of the prediction $\alpha$ weights per class for clinical condition prediction.}
\begin{center}
\resizebox{0.9\linewidth}{!}{%
\begin{tabular}{lrrrrrrr}
\toprule
\textbf{Class} & \textbf{EHR} & \textbf{CXR} & \textbf{DN} & \textbf{RR} & \textbf{Missingness} & \textbf{Low} & \textbf{High} \\
\midrule
Acute Myocardial Infarction                & 0.312 & 0.131 & 0.211 & 0.241 & 0.026 & 0.040 & 0.039 \\
Congestive Heart Failure                   & 0.365 & 0.061 & 0.078 & 0.413 & 0.022 & 0.030 & 0.030 \\
Atrial Fibrillation                        & 0.271 & 0.074 & 0.394 & 0.131 & 0.031 & 0.050 & 0.050 \\
Pulmonary Embolism                         & 0.272 & 0.211 & 0.183 & 0.202 & 0.026 & 0.055 & 0.051 \\
Pneumonia                                  & 0.285 & 0.188 & 0.214 & 0.218 & 0.022 & 0.037 & 0.036 \\
Sepsis                                     & 0.219 & 0.183 & 0.205 & 0.230 & 0.043 & 0.060 & 0.059 \\
Acute Renal Failure                        & 0.307 & 0.135 & 0.079 & 0.339 & 0.038 & 0.052 & 0.051 \\
Chronic Kidney Disease                     & 0.217 & 0.170 & 0.069 & 0.436 & 0.030 & 0.040 & 0.039 \\
Liver Cirrhosis                            & 0.254 & 0.165 & 0.230 & 0.290 & 0.013 & 0.024 & 0.024 \\
Diabetes Mellitus                          & 0.257 & 0.187 & 0.175 & 0.275 & 0.022 & 0.043 & 0.042 \\
Hypertension                               & 0.575 & 0.107 & 0.111 & 0.083 & 0.033 & 0.046 & 0.045 \\
Stroke                                     & 0.494 & 0.165 & 0.074 & 0.110 & 0.045 & 0.058 & 0.054 \\
Chronic Obstructive Pulmonary Disease      & 0.254 & 0.273 & 0.177 & 0.186 & 0.025 & 0.044 & 0.042 \\
Asthma                                     & 0.356 & 0.343 & 0.095 & 0.123 & 0.020 & 0.032 & 0.031 \\
Deep Vein Thrombosis                       & 0.332 & 0.211 & 0.175 & 0.166 & 0.030 & 0.043 & 0.043 \\
Peripheral Arterial Disease                & 0.306 & 0.090 & 0.262 & 0.197 & 0.035 & 0.055 & 0.054 \\
Anemia                                     & 0.287 & 0.181 & 0.198 & 0.242 & 0.023 & 0.035 & 0.034 \\
Gastrointestinal Bleeding                  & 0.194 & 0.094 & 0.155 & 0.452 & 0.028 & 0.039 & 0.039 \\
Cancer                                     & 0.213 & 0.180 & 0.109 & 0.279 & 0.067 & 0.077 & 0.075 \\
Infection                                  & 0.362 & 0.129 & 0.123 & 0.196 & 0.055 & 0.068 & 0.067 \\
Inflammatory Disease                       & 0.218 & 0.115 & 0.051 & 0.469 & 0.043 & 0.053 & 0.052 \\
Neurological Disorder                      & 0.271 & 0.112 & 0.112 & 0.384 & 0.033 & 0.045 & 0.043 \\
Trauma                                     & 0.420 & 0.132 & 0.064 & 0.282 & 0.028 & 0.038 & 0.037 \\
Fracture                                   & 0.308 & 0.056 & 0.238 & 0.330 & 0.016 & 0.027 & 0.026 \\
Other                                      & 0.456 & 0.083 & 0.166 & 0.213 & 0.022 & 0.030 & 0.029 \\
\bottomrule
\end{tabular}}
\end{center}
\label{tab:pheno_alpha}
\end{table}

\newpage
\subsection{In-hospital Mortality Alpha Weights Comparison}

In Table~\ref{tab:inho_alpha} we present the learned modality weights for the In-hospital Mortality task. The distribution shows that while each modality contributes to the final prediction, the model strategically adjusts the weights across modalities and confidence levels in a data-driven manner. This analysis reinforces the notion that \texttt{MedPatch} not only integrates information at multiple stages but also dynamically prioritizes contributions based on the task and input characteristics.

\begin{table}[h]
\caption{Modality Weights for In-hospital Mortality.}
\begin{center}
\resizebox{0.8\linewidth}{!}{%
\begin{tabular}{lcccccc}
\toprule
\textbf{Setting}           & \textbf{EHR} & \textbf{CXR} & \textbf{RR} & \textbf{Missingness} & \textbf{Low} & \textbf{High} \\
\midrule
In-hospital Mortality      & 0.199      & 0.172      & 0.180      & 0.163               & 0.141      & 0.146 \\
\bottomrule
\end{tabular}}
\end{center}
\label{tab:inho_alpha}
\end{table}

\subsection{Parameter Counts}
MedPatch outperforms or matches the top baselines in all settings and is between forty to ninety times lighter than the other top performing models when it comes to trainable parameters. The only notable exception is early fusion, where the parameter count is less, but so is its performance across most settings (it is comparable to MedPatch in bimodal phenotyping.) We summarize these results in Table \ref{parameters}.

\begin{table}[h]
\caption{Trainable parameters across different models for in-hospital mortality prediction and clinical phenotyping tasks.}
\begin{center}
\resizebox{0.6\linewidth}{!}{%
\begin{tabular}{lcc}
\toprule
\textbf{Model} & \textbf{Parameters (Mortality)} & \textbf{Parameters (Clinical Conditions)} \\
\midrule
\multicolumn{3}{l}{\textit{Unimodal}} \\
\midrule
EHR (LSTM) & 1.2M & 1.2M \\
CXR (ViT) & 27.6M & 27.6M \\
RR (BioBERT) & 0.4M & 0.4M \\
DN (BioBERT) & -- & 0.4M \\
\midrule
\multicolumn{3}{l}{\textit{Bimodal (EHR + CXR)}} \\
\midrule
Early  & 0.0M & 0.0M \\
Joint  & 28.8M & 28.9M \\
Late  & -- & -- \\
MedFuse & 23.9M & 23.9M \\
MeTra & 33.4M & 34.4M \\
\rowcolor[gray]{0.9} \texttt{MedPatch} & 0.4M & 1.8M \\
\midrule
\multicolumn{3}{l}{\textit{Multimodal}} \\
\midrule
Early & 0.0M & 0.1M \\
Joint & 29.2M & 29.7M \\
Late  & -- & -- \\
\rowcolor[gray]{0.9} \texttt{MedPatch} & 1.2M & 4.8M \\
\midrule
\multicolumn{3}{l}{\textit{Ensembles}} \\
\midrule
Avg. preds (EJL) & -- & -- \\
Avg. preds (Early only) & -- & -- \\
Avg. preds (Joint only) & -- & -- \\
Avg. preds (Late only) & -- & -- \\
\bottomrule
\end{tabular}%
}
\label{parameters}
\end{center}
\end{table}

\subsection{Alternative Patching Strategy}
We experiment with entropy-based patching instead of confidence patching and report the results in table \ref{entropy}. For probability $p_i$ returned by the pretrained confidence classifiers, we define entropy H as: \[
H(P) = -\sum_{i=1}^{n} p_i \log_2 p_i
\]

\begin{table}[ht]
\centering
\caption{Performance for Entropy Patching across mortality prediction and clinical phenotyping tasks.}
\begin{tabular}{l|cc|cc}
\hline
\textbf{Model} & \multicolumn{2}{c|}{\textbf{In-hospital Mortality}} & \multicolumn{2}{c}{\textbf{Clinical Conditions}} \\
               & \textbf{AUROC} & \textbf{AUPRC} & \textbf{AUROC} & \textbf{AUPRC} \\
\hline
Bimodal        & 0.867          & 0.543          & 0.774          & 0.439 \\
Multimodal     & 0.876          & 0.558          & 0.864          & 0.620 \\
\hline
\end{tabular}
\label{entropy}
\end{table}

\subsection{Non-Clinical Baselines}
To further demonstrate the effectiveness of our architecture, we compared MedPatch to another state-of-the-art baseline: HealNet \citep{hemker_healnet_2024}. Notably, HealNet was not specifically designed for ICU-related tasks, making it a suitable candidate for evaluating the generalizability of our approach. The original implementation of HealNet yielded subpar performance, prompting us to integrate our own pretrained encoders into the model. Although this modification led to improved results, it also introduced a substantial increase in the number of trainable parameters. All HealNet results are presented in Table \ref{healnet}.

\begin{table}[ht]
\centering
\caption{Performance and trainable parameters for HealNet variants across mortality prediction and clinical phenotyping tasks.}
\resizebox{\textwidth}{!}{%
\begin{tabular}{l|ccc|ccc}
\hline
\textbf{Variant} & \multicolumn{3}{c|}{\textbf{In-hospital Mortality}} & \multicolumn{3}{c}{\textbf{Clinical Conditions}} \\
                 & \textbf{AUROC} & \textbf{AUPRC} & \textbf{Parameters}   & \textbf{AUROC} & \textbf{AUPRC} & \textbf{Parameters} \\
\hline
HealNet Bimodal                         & 0.545 & 0.151 & 32.4M & 0.532 & 0.204 & 40.4M \\
HealNet + Early Hybrid Bimodal         & 0.863 & 0.526 & 35M   & 0.772 & 0.439 & 35M   \\
\textbf{HealNet + Early Hybrid Multimodal} & \textbf{0.875} & \textbf{0.550} & \textbf{38.8M} & \textbf{0.886} & \textbf{0.659} & \textbf{42.6M} \\
\hline
\end{tabular}
}
\label{healnet}
\end{table}

\newpage
\subsection{Statistical Significance Testing}
We conduct statistical significance testing for the results in table \ref{tab:bimodal} using the t-test adjusted for multiple comparisons. The results in table \ref{significane} confirm that differences among models are statistically significant. 
\begin{table}[h]
\centering
\caption{Paired $t$-tests on bootstrap replicates comparing \texttt{MedPatch} to other models. Negative $t$ indicates that \texttt{MedPatch} outperforms the comparator.}
\resizebox{\textwidth}{!}{%
\begin{tabular}{lrrrrrrrr}
\toprule
& \multicolumn{4}{c}{\textbf{Mortality}} & \multicolumn{4}{c}{\textbf{Phenotyping}} \\
\cmidrule(lr){2-5}\cmidrule(lr){6-9}
\textbf{Comparator} & \textbf{AUROC $t$} & \textbf{AUROC $p$} & \textbf{AUPRC $t$} & \textbf{AUPRC $p$} & \textbf{AUROC $t$} & \textbf{AUROC $p$} & \textbf{AUPRC $t$} & \textbf{AUPRC $p$} \\
\midrule
late  & $-156.18$ & $0.0$ & $-112.19$ & $0.0$ & $-2667.91$ & $0.0$ & $-2362.15$ & $0.0$ \\
joint & $-72.54$  & $0.0$ & $-49.93$  & $9.19 \times 10^{-274}$ & $-205.53$ & $0.0$ & $-148.07$ & $0.0$ \\
early & $-129.51$ & $0.0$ & $-96.16$  & $0.0$ & $-135.11$ & $0.0$ & $-15.90$  & $6.48 \times 10^{-51}$ \\
metra & $-31.37$  & $6.17 \times 10^{-151}$ & $-70.76$ & $0.0$ & $-240.99$ & $0.0$ & $-271.88$ & $0.0$ \\
\bottomrule
\end{tabular}
}
\label{significane}
\end{table}

\newpage
\subsection{Beta Weights Analysis}
In this section, we analyze the learned beta weights that are used to merge the different loss components during training. These weights determine the importance that \texttt{MedPatch} assigns to each branch of the architecture—namely, the late, high-confidence, and low-confidence modules—when computing the overall loss function.

Table~\ref{tab:weights_separate} provides a comparison of the weight metrics under bimodal and multimodal settings for both the In-hospital Mortality and Phenotyping tasks. From these results, we can observe a marked difference in how the model distributes weight between the components depending on the task and the number of available modalities. For example, in the bimodal In-hospital Mortality configuration the late fusion branch receives a dominant weight (0.910), indicating that it contributes the most to the final prediction error. In contrast, when additional modalities are incorporated, the weights are more evenly distributed (e.g., the late weight drops to 0.701 in the multimodal setting) as the model leverages the complementary strength of the high- and low-confidence predictions. This adaptive weighting underscores the dynamic nature of our fusion strategy and its capacity to assign importance based on the input structure.

\begin{table}[h]
\caption{Comparison of Weight metrics in \texttt{MedPatch}'s bimodal and multimodal settings, split by In-hospital Mortality and Phenotyping tasks. The bold results are the highest per weight group (Late, High, Low) per row.}
\begin{center}
\begin{tabular}{l ccc}
\toprule
{\textbf{Setting}} 
    & \multicolumn{3}{c}{\textbf{Weights}} \\
\cmidrule(lr){2-4}
    & \textbf{Late} & \textbf{High} & \textbf{Low} \\
\midrule
\multicolumn{4}{c}{\textbf{In-hospital Mortality}}\\
\midrule
Bimodal    & \textbf{0.910} & 0.062 & 0.028 \\
Multimodal & \textbf{0.701} & 0.187 & 0.112 \\
\midrule
\multicolumn{4}{c}{\textbf{Phenotyping}}\\
\midrule
Bimodal    & \textbf{1.00}  & 0.000 & 0.000 \\
Multimodal & \textbf{0.999} & 0.001 & 0.000 \\
\bottomrule
\end{tabular}
\end{center}
\label{tab:weights_separate}
\end{table}

\end{document}